\documentclass[aps,prb,twocolumn,showpacs,floatfix]{revtex4}
\pdfoutput=1
\usepackage{graphicx}
\usepackage{amssymb}
\usepackage{amsfonts}
\usepackage{amsmath}
\usepackage{dcolumn}

\usepackage{makecell}

\usepackage{color}
\usepackage{stackrel}
\usepackage{latexsym}

\usepackage[normalem]{ulem}
\usepackage{times}
\usepackage{bm}
\usepackage{hyperref}
\usepackage{accents}

\begin{document}
\title{Properties of 2D and Quasi-2D Dipolar Bosons with Non-zero Tilt Angles at T=0 }

\author{Pengtao Shen}
\affiliation{Department of Physics, Kent State University, Kent , OH 44242, USA}
\author{Khandker F. Quader}
\affiliation{Department of Physics, Kent State University, Kent , OH 44242, USA}

\date{\today}

\begin{abstract}
Recent experimental advances in creating stable dipolar bosonic systems, including polar molecules with large electric
dipole moments, have led to vigorous theoretical activities. Recent reporting of observation of roton feature in dipolar erbium has
provided added impetus to theoretical and experimental work. Here we discuss our mean-field theory work on 
2D and quasi-2D dipolar bosons with dipoles oriented at an angle to the direction perpendicular to the 
confining 2D plane, i.e. for {\it non-zero tilt angles}. Using Bogoliubov-de Gennes equations, we present results on a number of T=0 properties of both 2D and quasi-2D systems, such as excitation spectra, structure functions, sound velocities, quantum depletion, etc. We explore instabilities at varying tilt angle, density and dipolar coupling. We map out phase diagrams as  a function of tilt angle, dipole strength and density. We find the development of maxon-roton behavior leading to roton instabilities at large densities for small tilt angles, and at low densities for large tilt angles. The behavior is anisotropic in k-space; accordingly the roton instabilities occur first in the $k_y$ direction, suggestive of inhomogeneity and stripe phase, with density mode becoming soft in the $y$-direction. Beyond a critical tilt angle, at any density, the dipolar system collapses owing to a phonon instability. We discuss similarities and differences between the properties of 2D and quasi-2D dipolar systems at non-zero tilt angles.
\end{abstract} 
  
 \pacs{67.85 -d,67.85.Jk, 67.85.De, 05.30.Rt, 5.30.Jp}

\maketitle

\section{Introduction}
The nature of excitations and phases of interacting Bose systems has been a subject of longstanding interest. 
It has been known for a long time~\cite{feynman-rmp1957} that the excitation spectrum~\cite{he4-expt1971} in $^4He$  displays a linear-k phonon mode at low momentum that goes over in a continuous fashion to a a parabolic-like behavior with a minimum at finite momentum called the "roton". Since then, 
there has been interest in finding out if, in addition to  the low-momentum (k) linear-k "phonon" mode,  a roton feature could be found in other atomic Bose systems.  The extraordinary development of the field of ultracold atoms, tremendously advanced 
by novel experimental techniques, has, over the past several years, led to intense research along these lines.
The maxon-roton feature in $^4He$ is driven by strong inter-atomic forces. By comparison, the interatomic forces in dilute ultracold Bose gases 
are weak and contact interactions in BECs do not result in roton-like features in excitation spectra of these gases. 

In recent years, systems with long-range and anisotropic interaction have generated considerable interest. Among these are dipolar interactions between bosonic atoms or polar molecules. Experimentally, dipolar BECs have been realized in chromium~\cite{nature-dg-stable} ($^{52}Cr$), and in lanthanide atoms (such as dysprosium, erbium~\cite{Ferlaino2012}), which have larger magnetic moments. Recent reporting~\cite{ferlaino-roton2018} of observation for the first time of roton mode in dipolar $^{166}Er$ in cigar-shape trap geometry constitute a significant development. Realization of high phase-space density  systems  of polar molecules, such as,  $^{87} Rb^{133}Cs$~\cite{molony2014creation,takekoshi2014ultracold}, $^{41}K^{87}Rb$~\cite{aikawa2009,aikawa2010} hold promise for realization of quantum degeneracy and 
dipolar BEC. In general, the electric dipole moments of the polar molecules are substantially larger than the magnetic dipole moments of atoms; e.g. 
RbCS system has sizable electric dipole moment $\sim$ 1.28 Debye. 

The long-range and anisotropic nature, and a region of attraction of dipolar interaction can give rise to novel quantum phases, even in dilute systems. 
Earlier mean-field theory work~\cite{santos2003}  had predicted the existence of roton mode in BECs with dipolar interactions. Subsequent microscopic 
calculations~\cite{mazzanti2009dynamics} using a combination of diffusion Monte Carlo and correlated basis function methods, also found a roton feature in the excitation spectrum. 
These and other theory work were in systems  where the dipoles were oriented perpendicular to the 2D plane.
Since applied electric field can be used to fix the orientation of 2D electric dipoles in a system, one of the 
degrees of freedom in these systems is the {\it tilt angle} fixed by the direction of the applied electric field. Much of the theory work on quasi 2D and other 2D Bose systems has been done for zero tilt angle,~\cite{Fischer,ticknor2011anisotropic} except for Quantum Monte Carlo (QMC) calculations~\cite{macia2012excitations,macia2014phase} on 2D bosons, and some work on quasi-2D system~\cite{fedorov2014two}.

In this paper, we focus on obtaining properties of homogeneous 2D and quasi-2D Bose gas at zero temperature (T=0)
for {\it arbitrary tilt angle} $\theta$ at the mean-field level. 
We report here the results of our calculations of a number of properties of these two 2D systems, namely the excitation spectra, including roton-maxon features, structure functions, instabilities and phase diagram, roton position, sound velocities, Landau critical velocity, and quantum depletion. 
We compare our results for 2D and quasi-2D cases, and comment on how results from mean-field calculation, such as this, compare with findings 
of QMC calculations.

Our calculations allow us to explore instabilities at varying tilt angle, density and dipolar coupling in both 2D and quasi-2D systems 
with dipolar interaction. We map out phase diagrams as  a function of tilt angle, dipole strength and density. We find the development of maxon-roton behavior leading to roton instabilities at large densities for small tilt angles, and at low densities for large tilt angles. The behavior is anisotropic in k-space; accordingly the roton instabilities occur first in the $k_y$ direction, with density mode becoming soft in the $y$-direction. This may be suggestive of inhomogeneity and a stripe phase. Beyond a critical tilt angle, at any density, the dipolar system collapses owing to a phonon instability. We also calculate and discuss
the issue of quantum depletion in these systems. Overall, the sets of results are aimed at providing a comprehensive picture of how the various properties of these 2D boson systems are affected by the breaking of rotational invariance through a non-zero tilt angle.

To provide a background on 2D Bose systems and the methods used, we begin with a quick review of Bogoliubov-de Gennes (BdG) theory in Sec IIA, and interactions between bosons in 2D in Sec. IIB. In Sec IIC, we discuss, in some details, dipolar interactions in homogeneous 2D and quasi-2D Bose systems, the 
BdG equations and the resulting energy spectra and instabilities. In Sec. III and IV we present various results for 2D and quasi-2D systems respectively.
We end with discussions in Sec. V.

\section{Background and Methods}

\subsection{Bogoliubov-de Gennes Theory}

The grand canonical Hamiltonian of an interacting Bose gas is given by~\cite{griffin2002,griffin2009}
\begin{eqnarray}
\hat K \equiv \hat H\ - \mu \hat N=\int {dr} {{\hat {\psi}} ^\dag(r)}[ - \frac{{{\hbar^2\nabla ^2}}}{{2m}} + {U_{ex}}(r) - \mu ]\hat \psi(r)\nonumber\\
+ \frac{1}{2}\int {drdr'} \hat{\psi ^\dag }(r)\hat {\psi ^\dag }(r')V(r - r')\hat \psi(r)\hat \psi(r')
\end{eqnarray}
where $\hat \psi (r)$ is the Bose annihilation field operator. This is expanded as: $\hat \psi (r)={\phi _0}(r)+\tilde \psi (r)$, with  $\phi_0(r)=\langle \hat \psi (r) \rangle$ being the condensate wave function and $ \tilde\psi (r)=\hat \psi (r)-{\phi _0}(r)$  the fluctuation from the condensate.
Substituting the expansion into Eq. (1),  the grand canonical Hamiltonian $\hat K$ can be expanded in orders of $\tilde{\psi}$.
Then, retaining terms up to 2nd order in fluctuation, $\tilde{\psi}$, and minimizing the grand canonical Hamiltonian, one obtains
the generalized Gross-Pitaevski equation for the condensate wave function $\phi_0(r)$,
\begin{equation}
[-\frac{{{\nabla ^2}}}{{2m}} + {U_{ex}}(r)+\int dr'{\vert \phi_0(r')\vert}^2 V(r-r') ]\phi_0(r)=\mu\phi_0(r)
\end{equation}





The BdG Hamiltonian, $K_{\text{BDG}}$ is then given by:
\begin{eqnarray}
\hat K_{\text{BDG}}&=&\int {dr}  \phi ^{\ast}_0 (r) [-\frac{{{\hbar^2\nabla ^2}}}{{2m}} + {U_{ex}}(r) - \mu ]\phi_0(r)\nonumber\\
&+&\frac 1 2\int drdr'\phi^{\ast}_0(r)\phi_0^{\ast}(r')V(r - r')\phi_0(r)\phi_0(r')\nonumber\\
&+&\int dr {\tilde {\psi}} ^\dag(r) [ - \frac{{{\hbar^2\nabla ^2}}}{{2m}} + {U_{ex}}(r) - \mu ] {\tilde\psi}(r)  \nonumber\\
&+&\int drdr' {\vert \phi_0(r')\vert}^2 {{\tilde {\psi}} ^\dag(r)}  V(r-r'){\tilde\psi}(r) \nonumber\\
&+&\int drdr' \phi ^{\ast}_0 (r)\phi_0(r') {{\tilde {\psi}} ^\dag(r')}V(r-r') {\tilde\psi}(r)\nonumber\\
&+&\frac 1 2 \int drdr' {\tilde {\psi}} ^\dag(r){\tilde {\psi}} ^\dag(r') V(r-r')\phi_0(r) \phi_0(r') \nonumber\\
&+&\frac1 2\int drdr'\phi ^{\ast}_0 (r)\phi ^{\ast}_0 (r')V(r-r'){\tilde\psi}(r){\tilde\psi}(r')
\end{eqnarray}

In homogeneous case, $U_{ex}(r)=0$, the ground state is $\phi_0(r)=\phi_0=\sqrt n_0$ and  $\mu=\int dr'{\vert \phi_0(r')\vert}^2 V(r-r') ]=n_0V(0)$. Taking Fourier transform $\tilde {\psi}(r)=\frac 1 {\sqrt V}\sum_{k\neq 0} b_k e^{ikr}$, one obtains
\begin{eqnarray}
\hat K_{BDG}&=&-\frac{N_0^2}{2V} V(0)+\sum_{k\neq 0}[\hbar^2 k^2/{2m}+V(k) n_0]b^\dag_k b_k\nonumber\\
&+&\sum_{k\neq 0}\frac 1 2 V(q)n_0 (b^\dag_k b^\dag_{-k} +b_kb_{-k})
\end{eqnarray}

The Hamiltonian is then diagonalized by Bogoliubov transformation~\cite{pethick-smith2008} resulting in 
\begin{eqnarray}
\hat K=-\frac{N_0^2}{2V} V(0)+\sum_{k\neq0}\varepsilon(k)b^\dag_kb_k
\end{eqnarray}

where $\varepsilon(k)$ is the Bogoliubov excitation spectrum
\begin{eqnarray}
\varepsilon(k)=\sqrt{(\hbar^2 k^2/{2m})^2+n_0V(k) \hbar^2 k^2/{m}}
\end{eqnarray}

An equivalent result can be derived in the random phase approximation (RPA) approach. If in the Hartree approximation the density-density response is 
$\chi_0(k,w)$, then the RPA response function is given by: 
\begin{eqnarray}
\chi_{\text{RPA}}(k,w)=\frac{\chi_0(k,w)}{1-V_k\chi_0(k,w)}
\end{eqnarray}

For a free boson system with condensation, and $n_0 \approx n$,
\begin{eqnarray}
\chi_0(k,w)=\frac{n_0}{\hbar w-\epsilon_k}-\frac{n_0}{\hbar w+\epsilon_k}
\end{eqnarray}
where $n_0$ is the condensate density, and $\epsilon_k=\hbar^2k^2/2m$ is the free particle kinetic energy.
Substitute Eq. (8) into Eq. (7), one can obtain the density-density response function of the interacting boson system with a condensate, 

\begin{eqnarray}
\chi(k,w)=\frac{2n_0\epsilon_k}{\hbar^2w^2-\epsilon^2_k-2n_0V_k\epsilon_k}
\end{eqnarray}
The pole of the above equation,  $\hbar w=\sqrt{\epsilon^2_k+2n_0V_k\epsilon_k}$, is exactly the Bogoliubov spectrum we obtained in Eq.(8).

\subsection{Boson interactions in 2D}

For completeness and to put dipolar interactions between bosons in perspective, we first briefly review a few points regarding interactions between bosons in 2D.
For a  sufficiently dilute Bose gas, it is normally adequate to consider  binary collisions, which means there is no third particle involved in such process. 
One may then describe binary collision between particles using the Born approximation $f(k,k')=\int dr e^{-ikr}V(r)e^{ik'r}$. 
For strong interactions, it may be necessary to obtain the scattering amplitude from Lippmann-Schwinger equation.
The off-shell scattering amplitude is then given by $ f(k,k')=m/\hbar^2\int dr e^{-ikr}V(r)\phi_k(r)$ , where $\phi_k(r)$ is the wave function of the relative motion with momentum $k$ after scattering. In general, one can obtain the scattering amplitude by solving the Schroedinger equation of relative motion\cite{landau-qm}. The leading order term of the scattering amplitude at low energy is given by s-wave scattering, $f(k)=\frac{2\pi}{ln(\kappa/k)+i\pi/2}$ in 2D, where 
$\kappa$ depends on the behavior of  $V(r)$ at small distance. One may omit the imaginary part when $k/\kappa\ll1$. In a uniform Bose-condensed gas, replacing k with $k_c$ defined by $2\mu=\hbar^2k_c^2/2m$, the coupling constant, $g =\frac{2\pi\hbar^2}{m}\frac1{\textrm{ln}(\kappa/k_c)} $.
In contrast with 3D, the result depends on the boson gas density\cite{schick}. It may be noted that 
in mean field, the coupling constant is always positive in the dilute limit.  

The discussion above pertains strictly to 2D. Experimentally, low dimensional systems are realized by applying a trap potential in confining direction. A simple form of trap is a cylindrically symmetric harmonic potential $U(z,\rho)=1/2(w_zz^2+w_{\rho}\rho^2)$, where z is the axial direction and $\rho$ the radial direction. Changing the aspect ratio of the trap potential $\lambda=w_z/w_{\rho}$ gives different trap forms. The trap is pancake-shaped when $\lambda>1$  and cigar-shaped when  $\lambda<1$. 

A 2D system, under a trap in axial direction $U(z)=1/2(w_zz^2)$, has discrete energy levels in axial direction is $E_i=\hbar w_z(i+1/2)$. In the ultra cold condition, $k_BT\ll\hbar w_z$ and  mean field inter-particle interaction $ng\ll\hbar w_z$ (n is the gas density, and
g the coupling constant), the gas is frozen in the ground state zero oscillation level and  kinematically is 2D.  On the other hand, the gas in the trap would be in crossover between kinematically two and three-dimensional behavior when $k_BT,ng\backsim\hbar w_z$.
However, the scattering process is still different depending on the characteristic length of the trap, $l_z=\sqrt{\hbar/mw_z}$ and three dimensional scattering length $a_{3D}$\cite{Petrov}. The system can thus be classified into two regimes:  quasi-2D and true 2D. 
In quasi-2D regime, $l_z\gg a_{3D}$. Although the particles are frozen in gourd state in axial direction, scattering is not directly affected by axial confinement. In true 2D regime, the axial confinement is tightened further,  $l_z\backsim a_{3D}$. The tight confinement makes the scattering take place only in the radial direction\cite{Petrov}. 

\subsection{Dipolar Boson Interaction in 2D}

The discussion above on interaction between bosons are extended in non-trivial way by bosonic (as well as fermionic) atoms or molecules with magnetic or electric dipole moments, for which the interactions are necessarily long-range and anisotropic. Consequently, 
the type of trap plays a very important role in the stabilization of dipolar gas system\cite{nature-dg-stable}. For example when dipoles are aligned pointing axial direction, pancake-shaped trap leads to more repulsive interaction and tend to stabilize the gas, while cigar-shaped trap leads to more attractive interaction and tend to collapse the gas.

As sketched in Fig.1, a 2D system of dipoles ${\bf d}$ with arbitrary orientation is characterized by a "tilt" angle $\theta$ between 
${\bf d}$ and the z-axis. The dipoles are aligned along $\theta$ by, for example, an electric field ${\bf E}$ (in the case of electric dipoles);
the dipoles move in the x-y plane and the vector separating them bear an angle $\phi$ with respect to the x-axis. 

\begin{figure}[h]
\includegraphics[width=16pc]{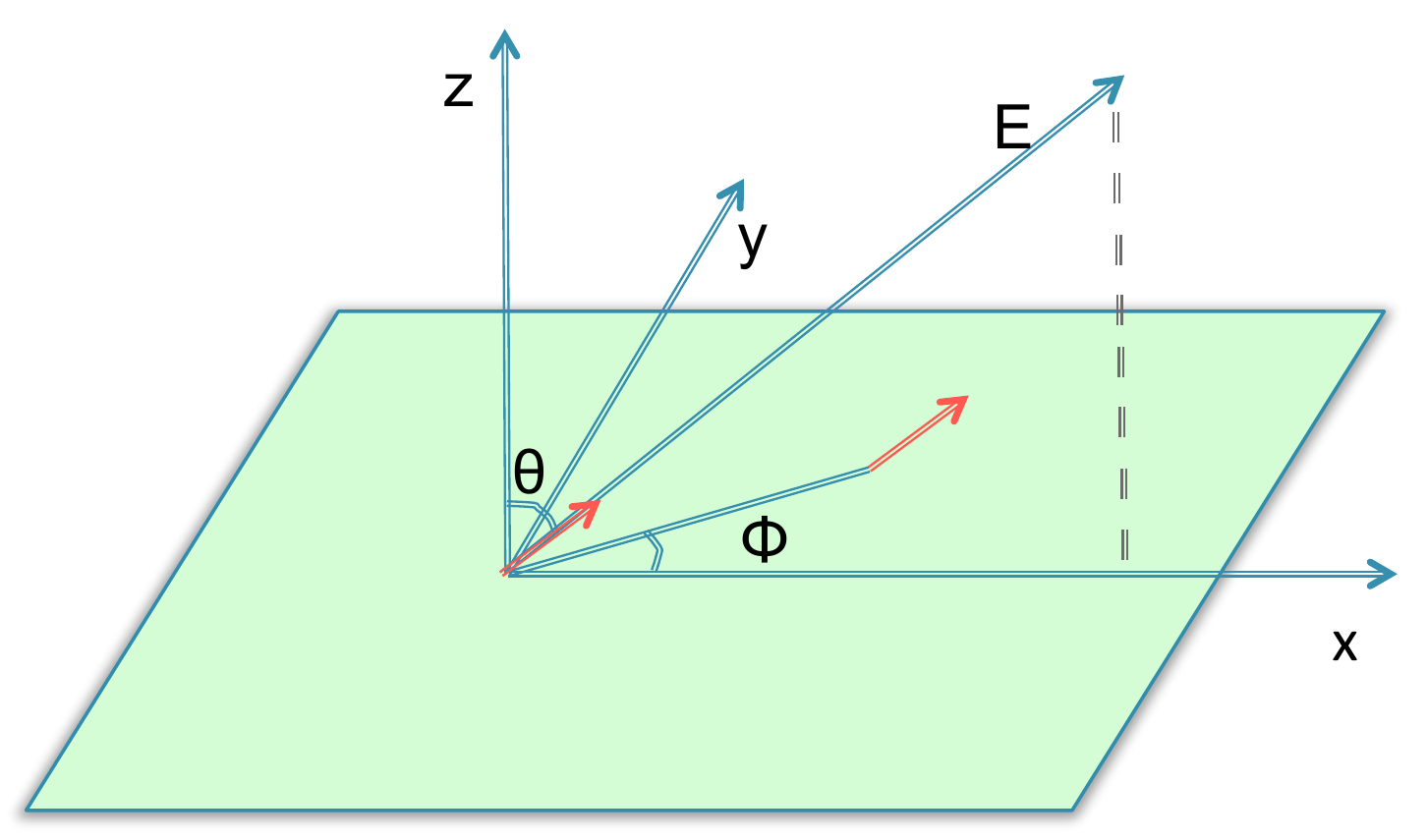}\hspace{4pc}%
\caption{(color online) 2D Dipoles in x-y plane with tilt angle $\theta$ that defines the direction of electric field E relative 
to the z-direction. $\Phi$ is the angle in x-y plane, relative to x-direction.}
\end{figure}

The 2D dipole-dipole interaction is given by,
\begin{eqnarray}
{V_{2D}}(r)&=& \frac{{{d^2}}}{{{r^3}}}(1 - 3{\sin ^2}\theta {\cos ^2}\phi ) \nonumber\\
&=& \frac{{{d^2}}}{{{r^3}}}[{P_2}(\cos \theta ) -\frac{3}{2}{\sin ^2}\theta \cos 2\phi ]
\end{eqnarray}
where  $d= |{\bf d}|$ is the strength of the dipolar interaction. The Fourier transform of 2D dipolar interaction in momentum space is:

\begin{eqnarray}
V(q)=\int^\infty_0d^2rV(r)e^{iqr}
\end{eqnarray}

The 2D integral however is ultraviolet divergent ($r\rightarrow 0$), and has to be regularized. As pointed out\cite{ddi-reg}, several regularization
schemes have been proposed in literature. Here, we shall consider two such regularizations resulting in (a) a homogeneous 2D system with a 
short-range cut-off $r_c$ in the Fourier transform; (b) a quasi-2D system wherein dipoles are confined to the lowest transverse sub-band
owing to an external harmonic  confining potential in the $z$-direction.~\cite{Fischer}
In both cases we consider arbitrary tilt angles.

(a) {\it Homogeneous 2D}: The choice of a value for the cut-off $r_c$ can be motivated in the case of polar molecules, at least, by taking $r_c$ to be of the order of the size of the molecule,
typically, of the order of several Bohr radius, $a_0$. 
Performing the integral  $\int^\infty_{r_c}$,  the 2D dipole-dipole interaction (DDI)  in momentum space can be written as $V(q)=V_s+V_l(q)$, with

\begin{eqnarray}
&V_s&=2\pi d^2 \frac{P_2(\cos\theta)}{r_c}\nonumber\\
&V_l(q)&=-2\pi d^2q({\cos^2}\theta-{\sin^2}\theta {\cos^2}\phi)
\end{eqnarray}
The short range part of the interaction $V_s$ depends on the tilt angle $\theta$: $V_s$ is positive at small $\theta$ and becomes negative at large $\theta$,
For low momentum transfer $q$, the long range term, $V_l(q)$,  depends linearly on $q$, and on tilt angle $\theta$ and azimuthal angle $\phi$. 
At zero tilt angle, $V_l(q)$ is isotropic and negative in all direction. For non-zero tilt angles, $V_l(q)$ becomes anisotropic: in $x$-direction, $V_l(q)$ gets less negative and becomes positive for $\theta>\pi/4$. In $y$-direction, it is always negative.
As will be seen, interesting consequences of 2D DDI, viz. instabilities of the system, happen when the DDI gets attractive in some direction. 
Hence, it is most interesting to consider the behavior in the $y$-direction, i.e. $\phi=\pi/2$, when the DDI is most attractive. Following standard practice, we define the dipolar interaction length $a_{dd}=\frac{\hbar^2d^2}m$. $a_{dd}$ can vary from $10a_0$ for magnetic dipoles to $10^4a_0$ for electric dipoles. 
On scaling with $a_{dd}$, we obtain

\begin{eqnarray}
V(q,\phi)=2\pi d^2 \left[\frac{P_2(\cos\theta)}{r_c}-q({\cos^2}\theta-{\sin^2}\theta {\cos^2}\phi)\right]\nonumber\\
=2\pi \frac{\hbar^2}{m} \left[\frac{a_{dd}}{r_c}P_2(\cos\theta)-q a_{dd}({\cos^2}\theta-{\sin^2}\theta {\cos^2}\phi)\right]
\end{eqnarray}

The interaction in the $y$-direction correspond to the in-plane angle $\phi=\pi/2$ above.

The first term in Eq. (13) depends explicitly on the cut-off $r_c$, and is short-range in nature; thus similar to a contact interaction,
sometimes added to the dipole-dipole interaction for bosons.
Thus,  having an additional a repulsive contact interaction will not qualitatively change the results. 


(b) {\it Quasi-2D}: Consider homogenous dipolar bosons in $x$-$y$ plane with tilt angle $\theta$; the  dipoles are confined by a harmonic trap in $z$-direction  $V(z) = \frac{1}{2}mw_z^2{z^2}$. 
By Eq(1), the 3D grand canonical Hamiltonian is

\begin{eqnarray}
\hat K \equiv \hat H\ - \mu \hat N=\int {dr} {{\hat {\psi}} ^\dag(r)}[ - \frac{{{\nabla ^2}}}{{2m}} + V(z) - \mu ]\hat \psi(r) \nonumber\\
+ \frac{1}{2}\int {drdr'} \hat{\psi ^\dag }(r)\hat {\psi ^\dag }(r')V(r - r')\hat \psi(r)\hat \psi(r')
\end{eqnarray}

The dipolar bosons can be considered to be frozen in the ground state of $V(z)$,with ground energy $\varepsilon_0=\hbar w_z/2$ in $z$ direction if $hw_z\gg\mu_{2D}$, where 
$\mu_{2D}$ is the 2D chemical potential, defined below. The condensate wave function is  $\phi_0=\phi_0(z)={(\frac{mw_z}{\pi\hbar})}^{\frac14}e^{-mw_zz^2/2\hbar}$ and fluctuation has the form of $\phi_0(z)\tilde\phi(r)$. This is different from the pancake geometry, where the system 
can be in several oscillator levels.

Integrating Eq. (14) over $z$, with $\hat{\psi}(r)=\phi_0(z)\hat\phi(\rho)$, one obtains
\begin{widetext}
\begin{equation}
\hat K \equiv \hat H\ - \mu \hat N=\int {d\rho} {{\hat {\psi}} ^\dag(\rho)}[ - \frac{{{\nabla^2_2}}}{{2m}} +\varepsilon_0- \mu ]\hat \psi(\rho)
+\frac{1}{2}\int {d\rho d\rho'} \hat{\psi ^\dag }(\rho)\hat {\psi ^\dag }(\rho')\hat \psi(\rho)\hat \psi(\rho')\int dzdz'V(r-r')\phi^2_0(z)\phi^2_0(z')
\end{equation}
\end{widetext}

The 2D chemical potential is then defined as $\mu_{2D} = \mu-\varepsilon_0$, and the effective 2D interaction as $V_{2D}(\rho-\rho') = \int dzdz'V(r-r')\phi^2_0(z)\phi^2_0(z')$ $\mu_{2D} = n_0V_{2D}(0)$. From this, an effective 2D Hamiltonian is obtained
\begin{eqnarray}
\hat K_{2D}=\int {d\rho} {{\hat {\psi}} ^\dag(\rho)}[-\frac{{{\nabla^2_2}}}{{2m}}- \mu_{2D}]\hat \psi(r)\nonumber\\
+ \frac{1}{2}\int {d\rho d\rho'} \hat{\psi ^\dag }(\rho)\hat {\psi ^\dag }(\rho')V_{2d}(\rho-\rho')\hat \psi(\rho)\hat \psi(\rho')
\end{eqnarray}

The Fourier transform of $V_{2d}(\rho-\rho')$ is 
\begin{eqnarray}
V_{2d}(k)&=&\pi d^2[\frac8{3l_z\sqrt{2\pi}}P_2(\cos\theta)\nonumber\\
&-&2({\cos^2}\theta-{\sin^2}\theta {\cos^2}\phi) F(k)]
\end{eqnarray}
where $F(k)=k\text{exp}[\frac{(kd_z)^2}2]\text{erfc}(\frac{kd_z}{\sqrt2})$

Eq.(12) and Eq.(17) has similar features since $F(k)\sim k$ at small value although $F(k)$ saturates at large k.  

\section{Results: Homogeneous 2D Dipolar Bosons}

\subsection{Spectrum and Stability}
Using  Eq.(12) in Eq.(6), we immediately obtain the BDG spectrum. On scaling with the dipolar strength $a_{dd}$, ${\tilde{\varepsilon}}=\varepsilon/\frac{\hbar^2}{ma^2_{dd}}$, ${\tilde{q}}=qa_{dd}$ and ${\tilde{n}}=n_0a^2_{dd}$, we obtain a spectrum with dimensionless energy and momentum, for non-zero tilt angle 
$\theta$:

\begin{equation}
\tilde{\varepsilon}(\tilde{q},\phi) =\nonumber
\end{equation}
\begin{equation}
\sqrt{\frac{{\tilde{q}^4}}{4}+2\pi {\tilde{q}}^2 {\tilde{n}}\left(\frac{a_{dd}}{r_c}P_2(\cos\theta)-{\tilde{q}}({\cos^2}\theta-{\sin^2}\theta\cos^2\phi)\right)} 
\end{equation}

The spectrum is anisotropic and the energy in $y$-direction, corresponding to $\phi=\pi/2$ , is lower than in any other direction in plane.
We explore the stability of the homogeneous 2D dipolar Bose gas by considering the above expression for the energy spectrum. It is useful to consider 
two sectors with respect to tilt angles $\theta$ :


{\bf a.} $\theta<\cos^{-1}{\frac1{\sqrt3}}$ (= 0.955): the short range interaction is positive and the long range part in $y$-direction is negative. The spectrum is always real and positive at small momentum; see Fig. 2. However,  at sufficiently large scaled density, ${\tilde{n}}$, a roton-maxon feature, reminiscent of that in 
$^4He$, develops in the spectrum. For even larger density, the roton spectra becomes imaginary in $y$-direction; this can be seen as ``holes" 
at symmetric $\pm q$ region in the 3D plot in Fig. 2; the ``holes" signifying that the energy $\epsilon$ is no longer real. 
The corresponding 2D plot depicts this feature as negative of energy squared for $\phi = \pi/2$, i.e. the $y$-direction. 
At the same time,  as can be clearly seen in the 2D plot, the spectra in the x-direction ($\phi = 0$) remain real, positive. 
This means the existence of density waves along one direction and not along that perpendicular to it.  This suggests a possible transition of the BEC phase to a stripe phase.

\begin{figure}[b]
\includegraphics[clip, width=1.6in]{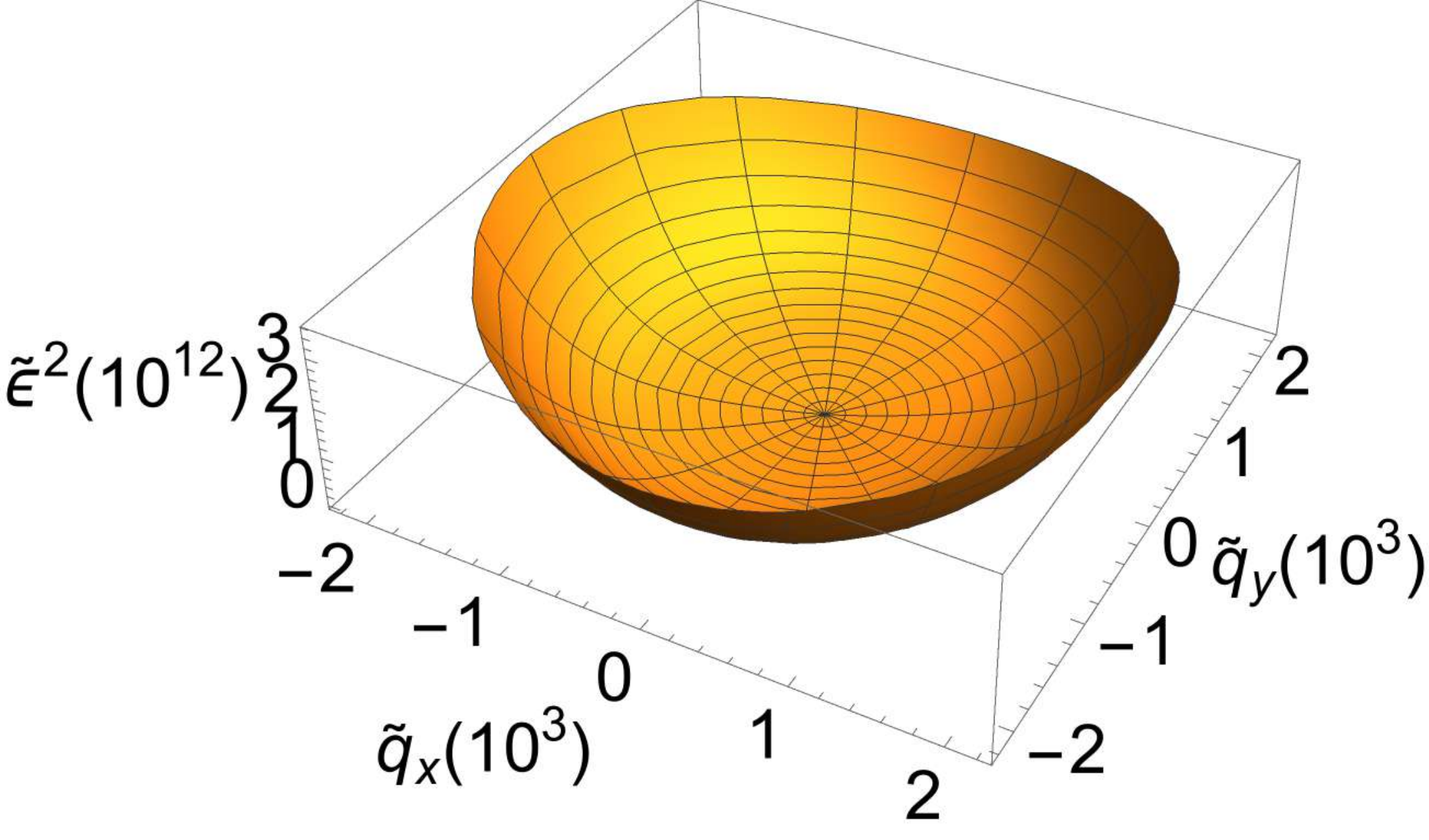}
\hspace{0.1in}
\includegraphics[clip, width=1.6in]{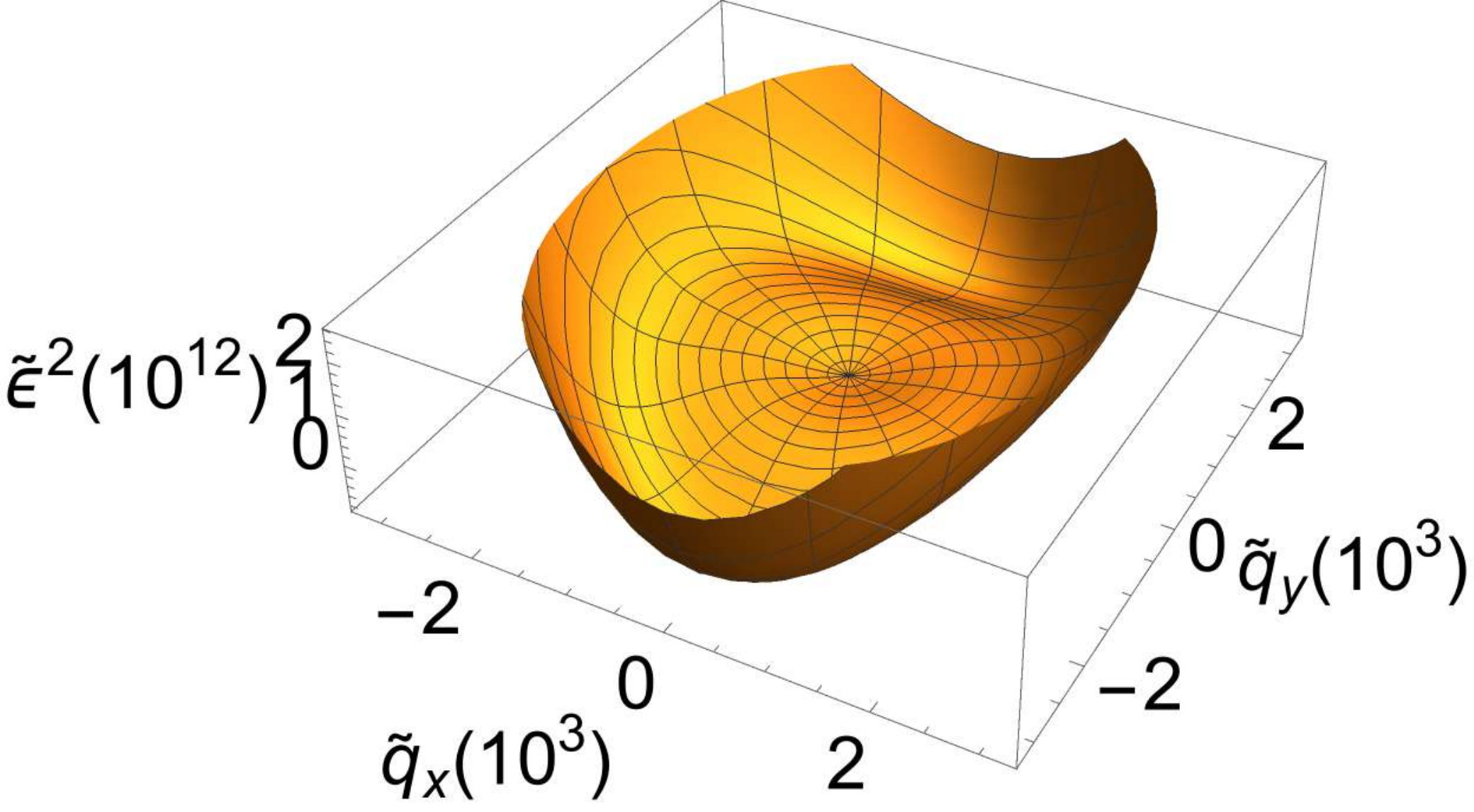}\\
\vspace{0.4in}
\includegraphics[clip, width=1.6in]{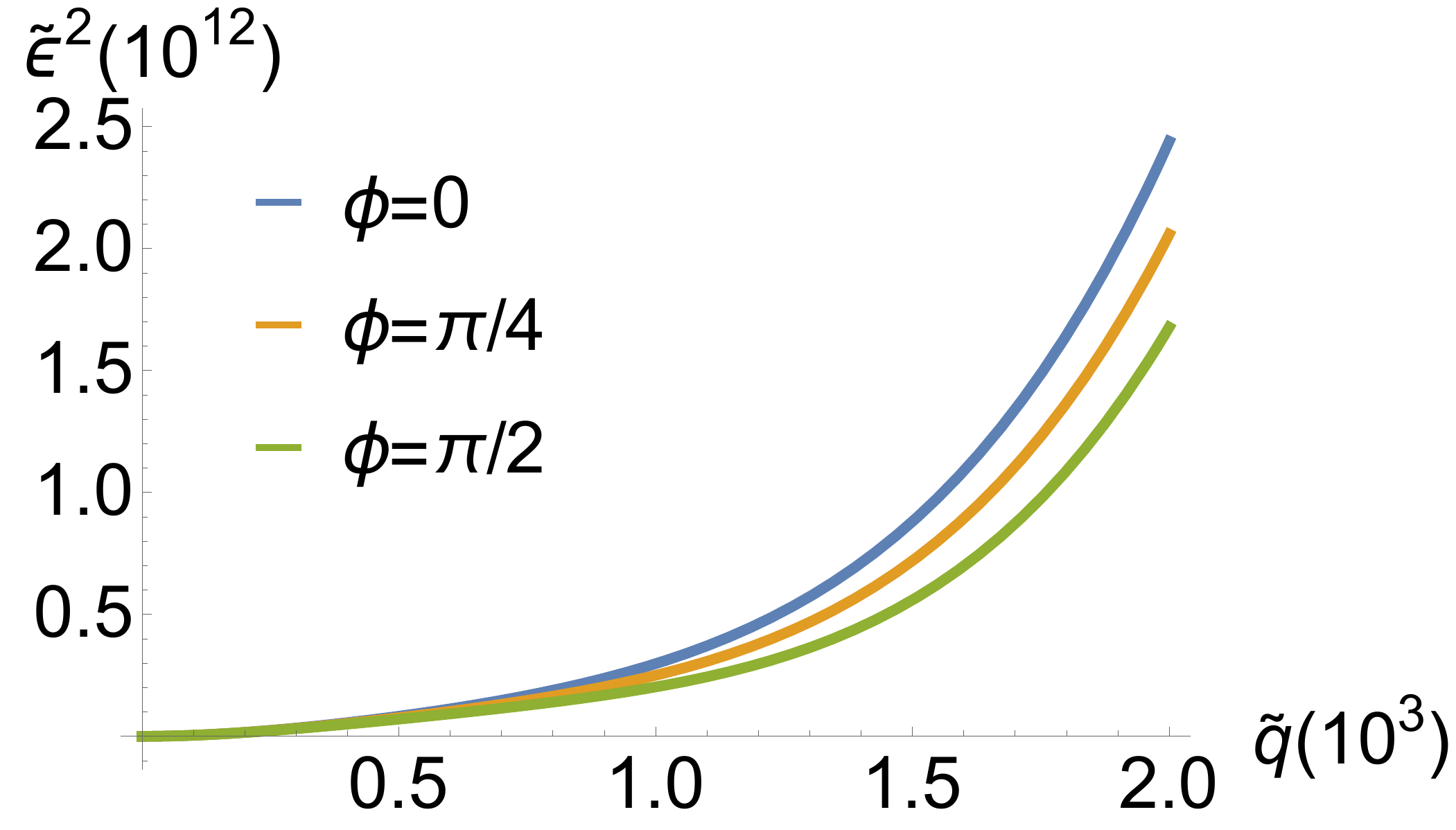}
\hspace{0.1in}
\includegraphics[clip, width=1.6in]{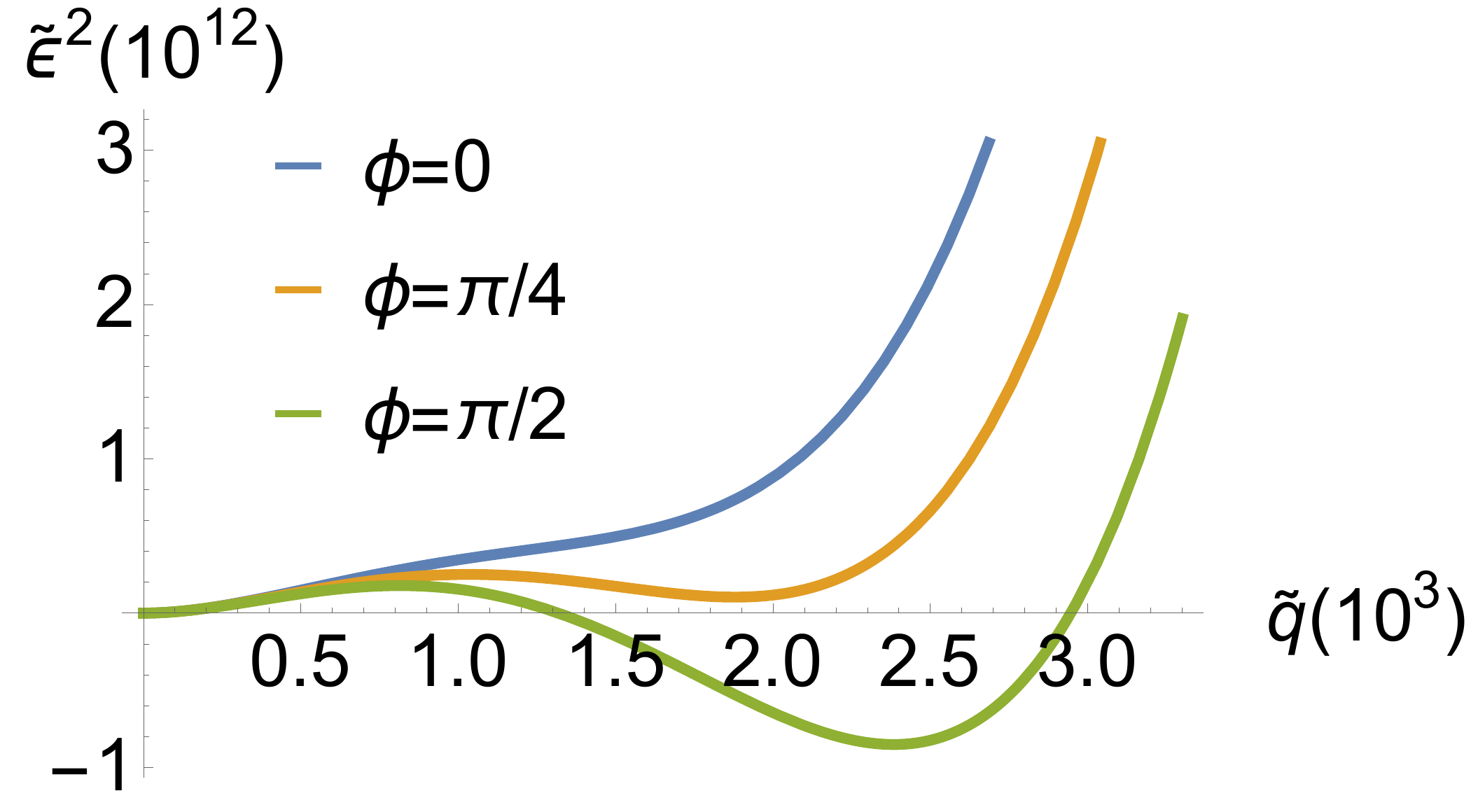}
\caption{(color online) Figure showing the spectra of dipolar Boson for fixed tilt angle $\theta=0.4$ at scaled density ${\tilde{n}}=100$ (left)
and  ${\tilde{n}}=200$ (right) respectively. The top 3D figures show the spectra in all directions $\phi$ on the x-y plane;
the bottom 2D figures show energy squared for selected values of the in-plane angle $\phi$ for the same two densities. 
$a_{dd}$ is taken to be $10^4a_0$, and $r_c=10a_0$.}
\end{figure}

\begin{figure*}
\includegraphics[clip, width=1.6in]{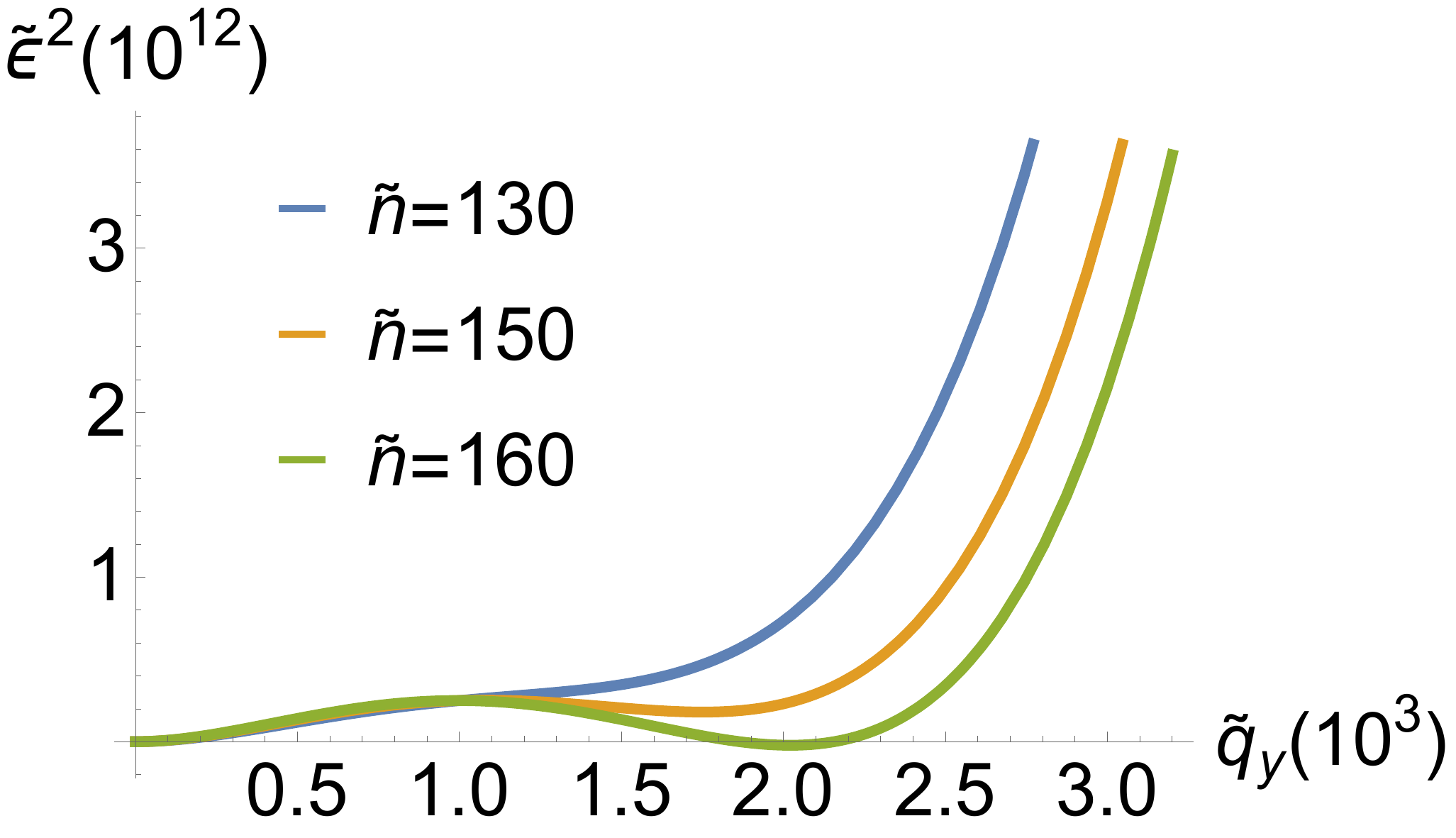}
\hspace{0.1in}
\includegraphics[clip, width=1.6in]{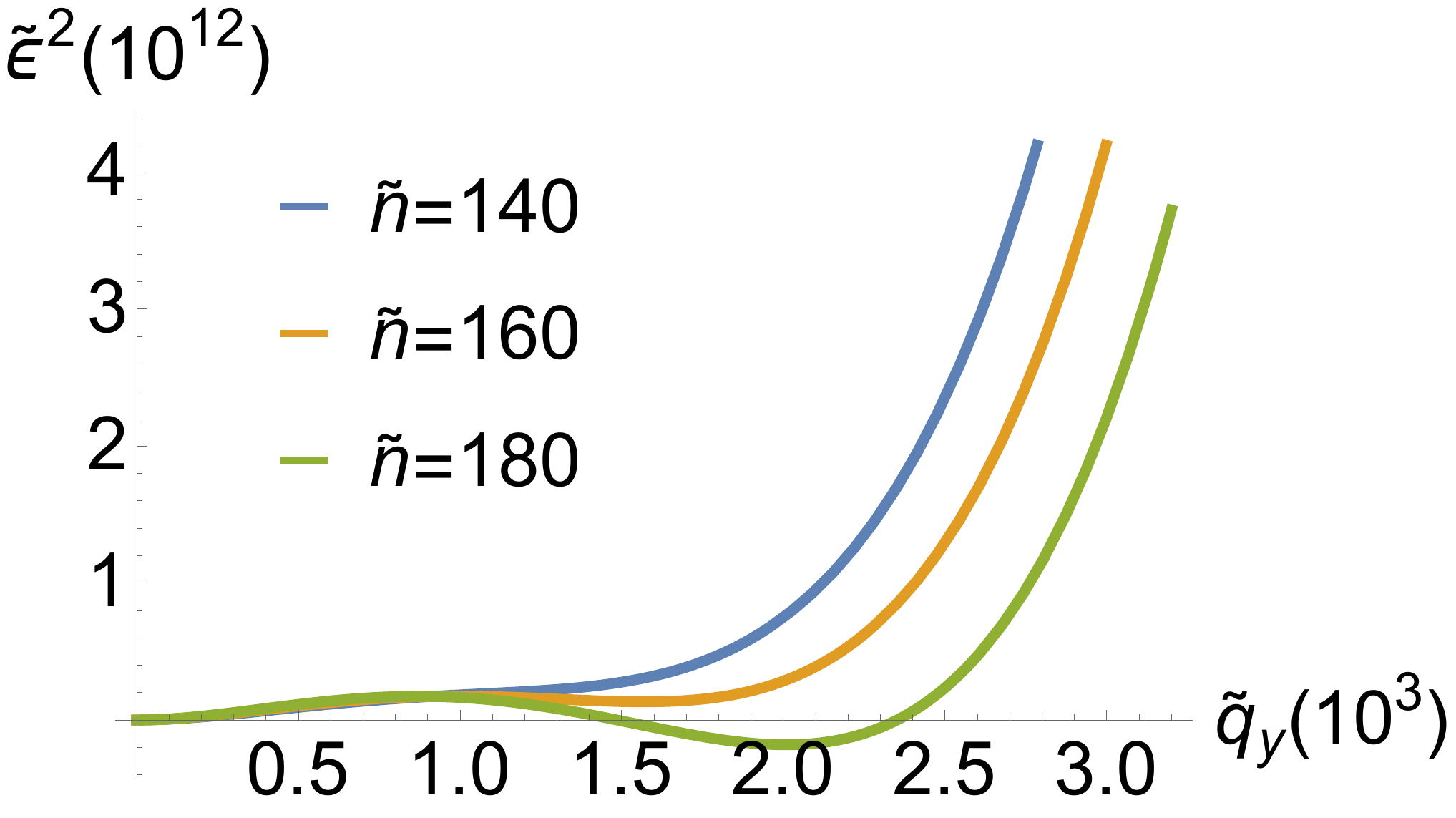}
\hspace{0.1in}
\includegraphics[clip, width=1.6in]{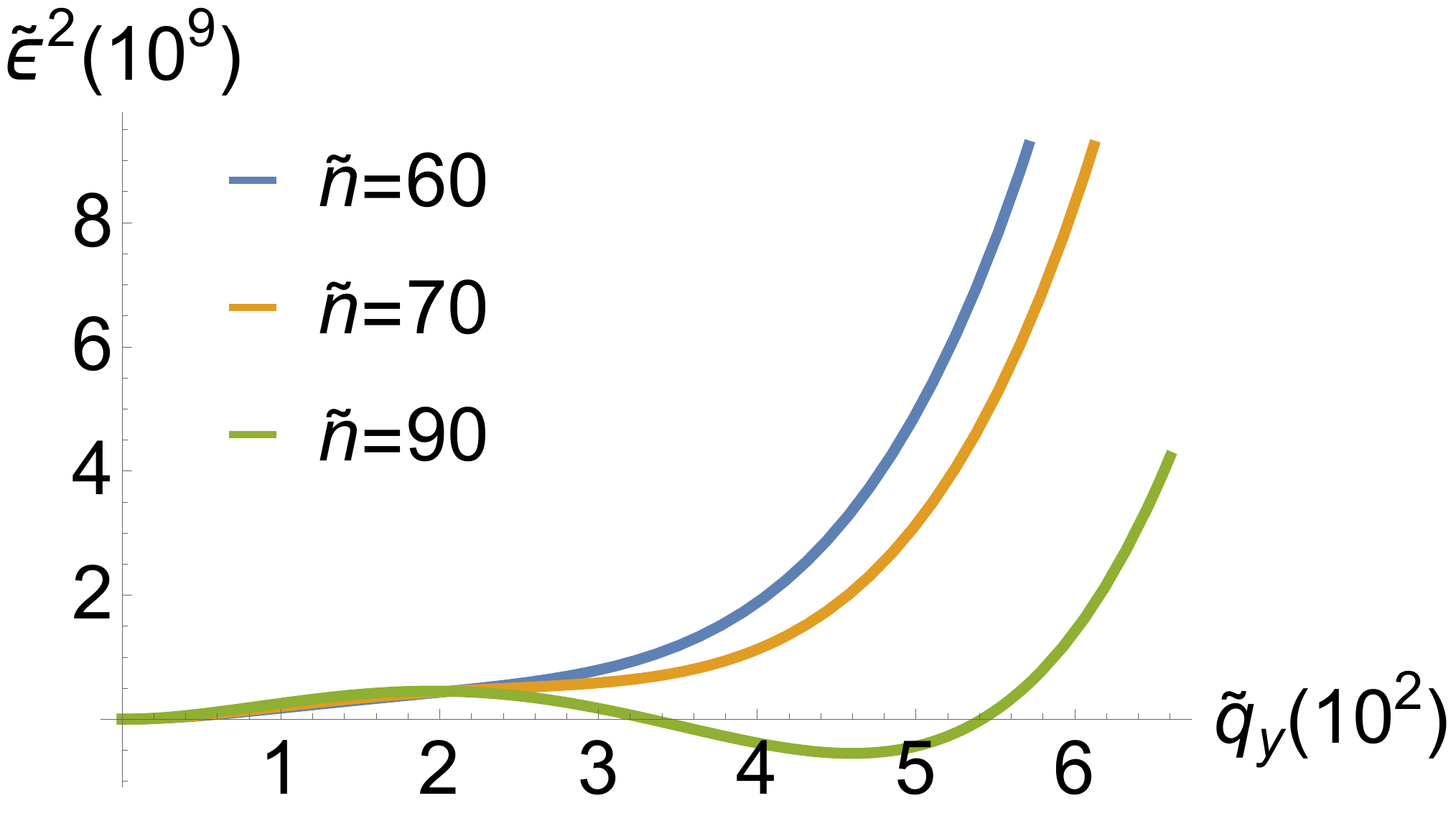}
\hspace{0.1in}
\includegraphics[clip, width=1.6in]{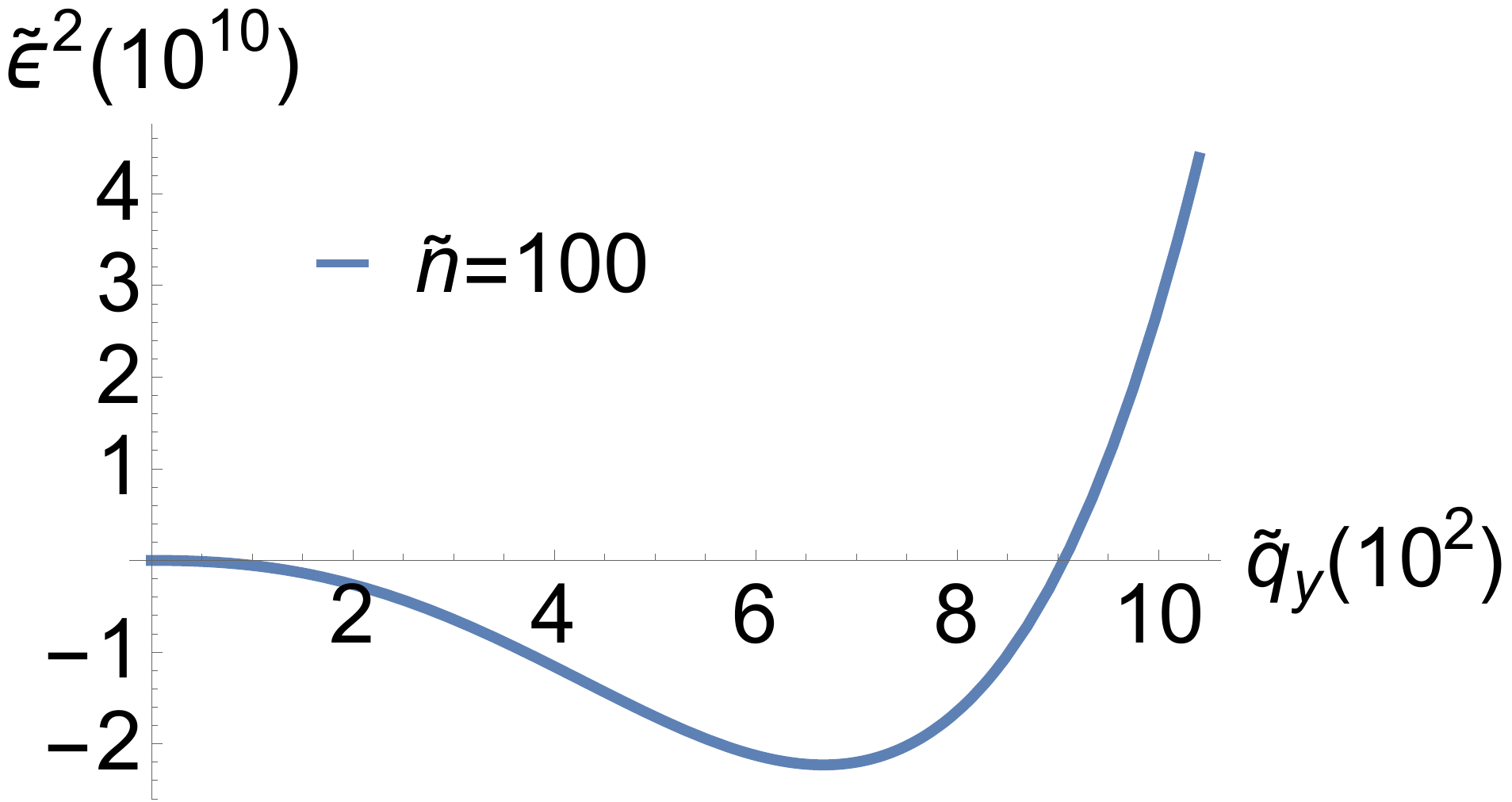}\\
\vspace{0.2in}
\includegraphics[clip, width=1.6in]{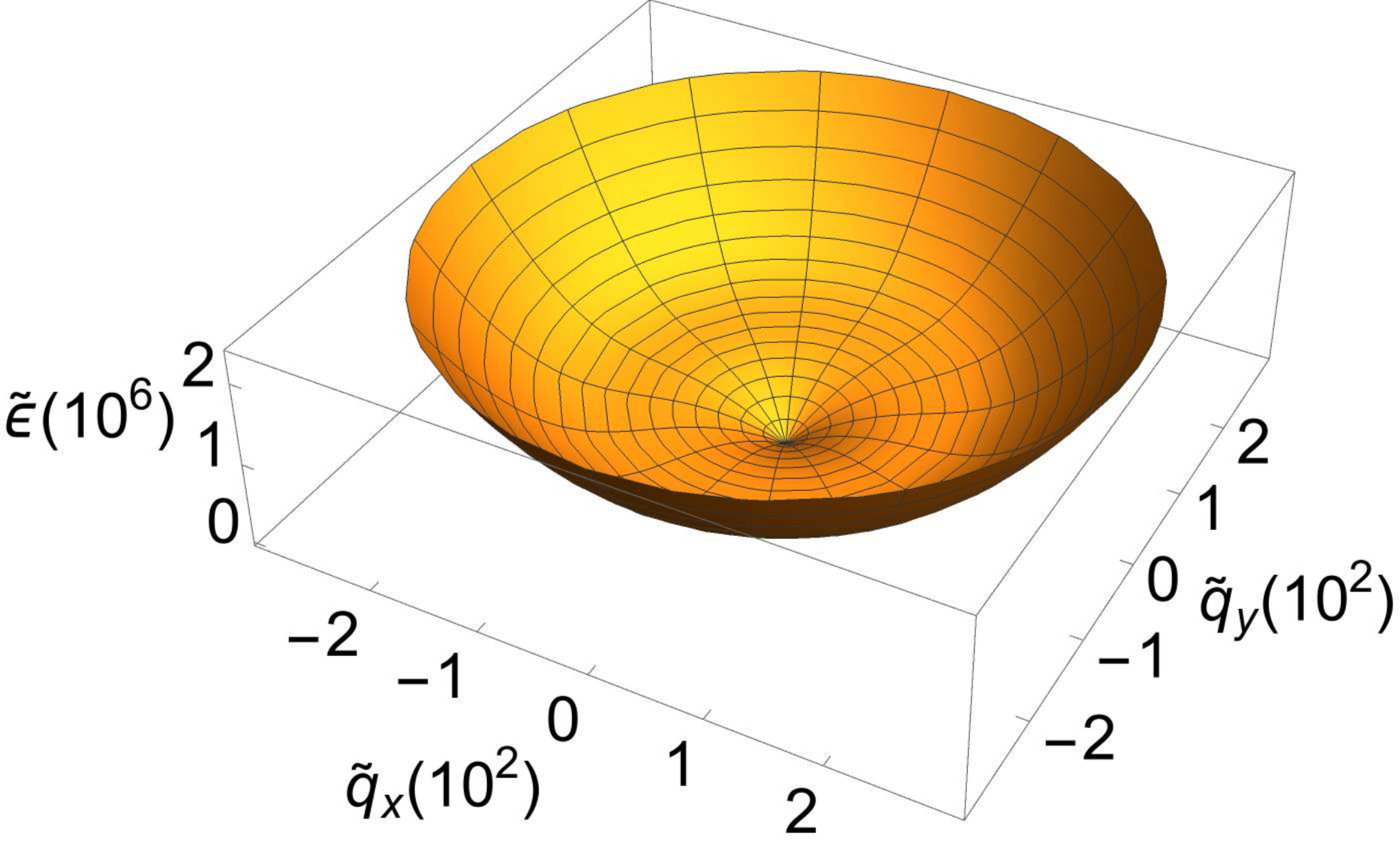}
\hspace{0.1in}
\includegraphics[clip, width=1.6in]{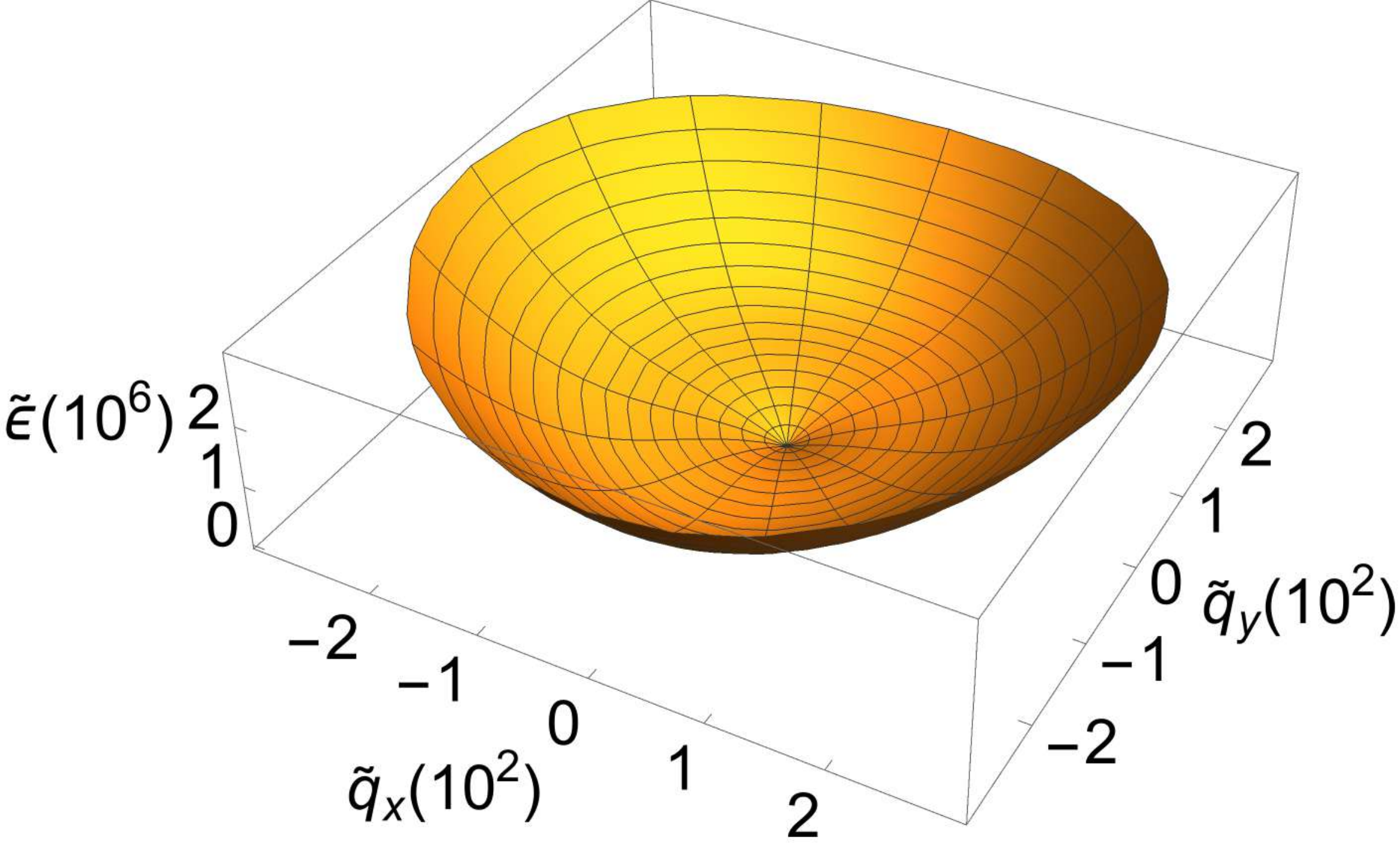}
\hspace{0.1in}
\includegraphics[clip, width=1.6in]{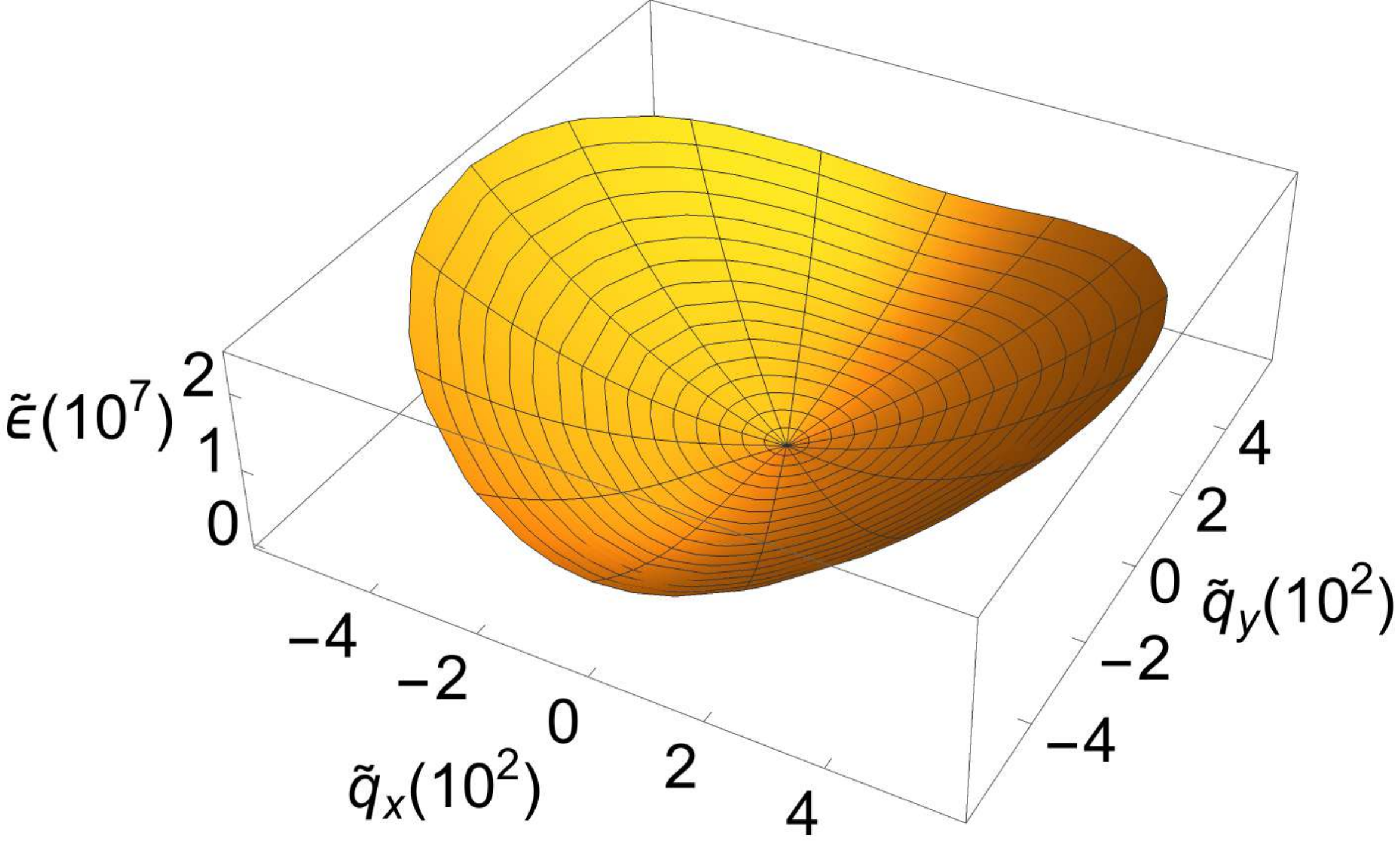}
\hspace{0.1in}
\includegraphics[clip, width=1.6in]{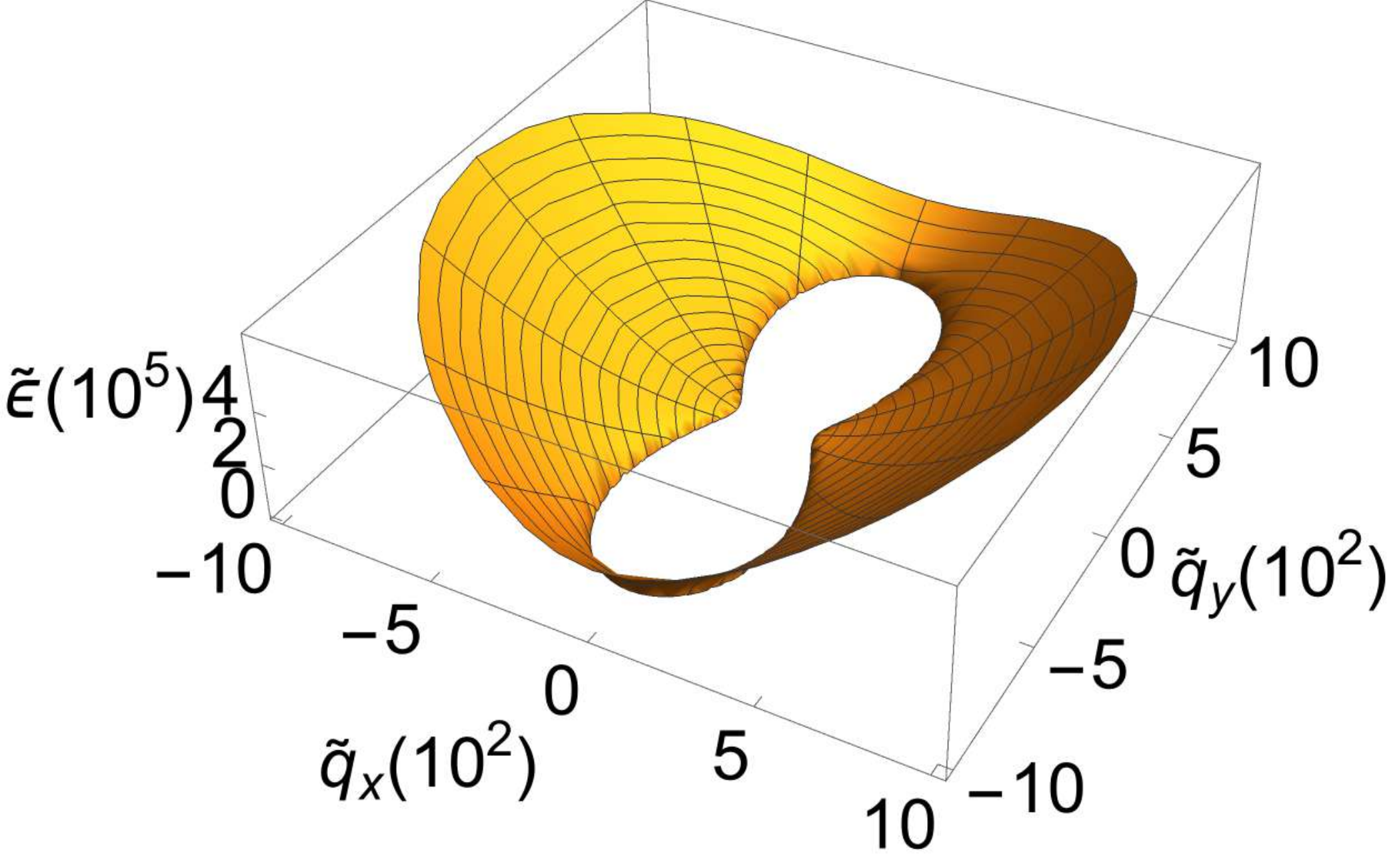}\\
\vspace{0.2in}
\includegraphics[clip, width=1.6in]{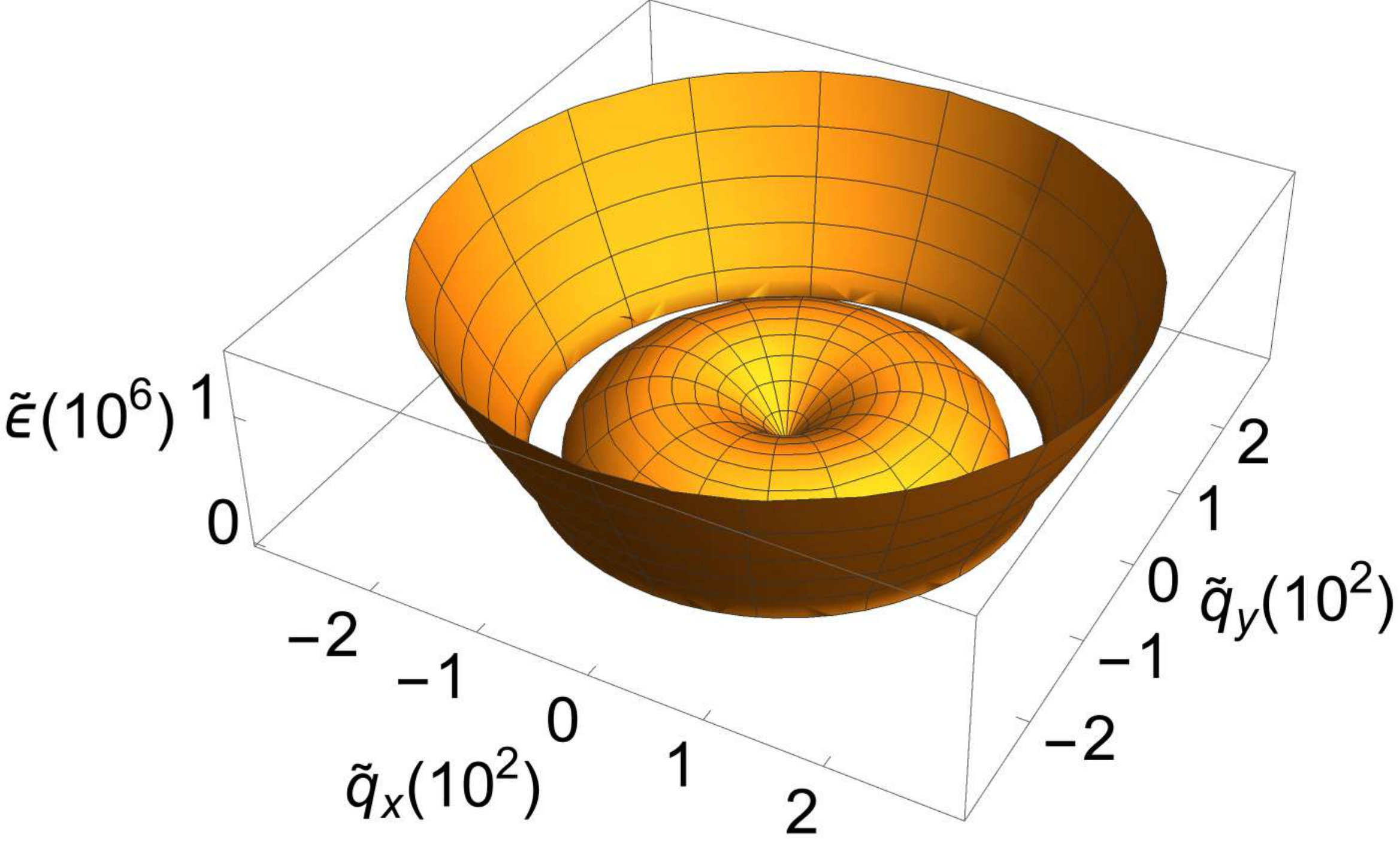}
\hspace{0.1in}
\includegraphics[clip, width=1.6in]{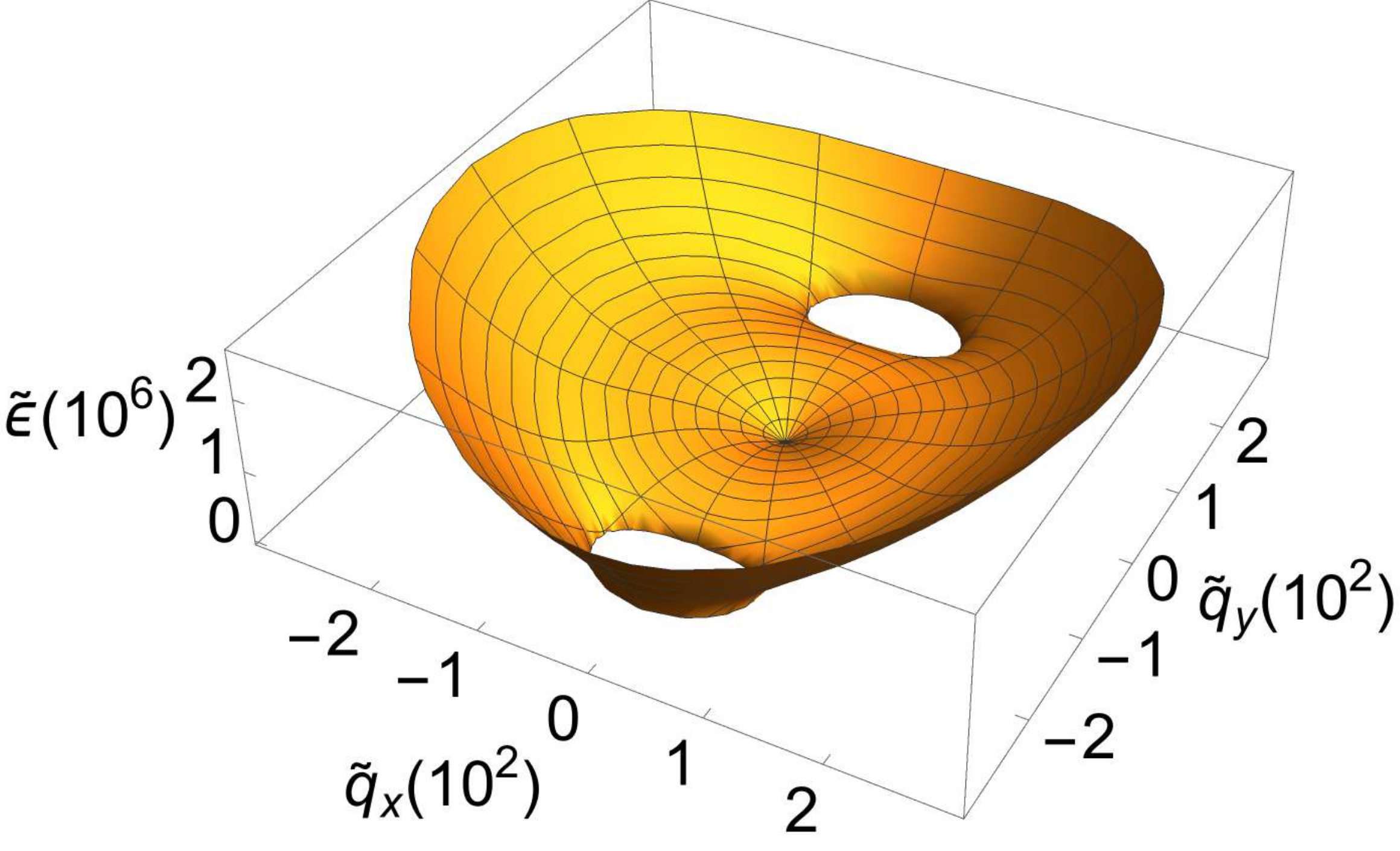}
\hspace{0.1in}
\includegraphics[clip, width=1.6in]{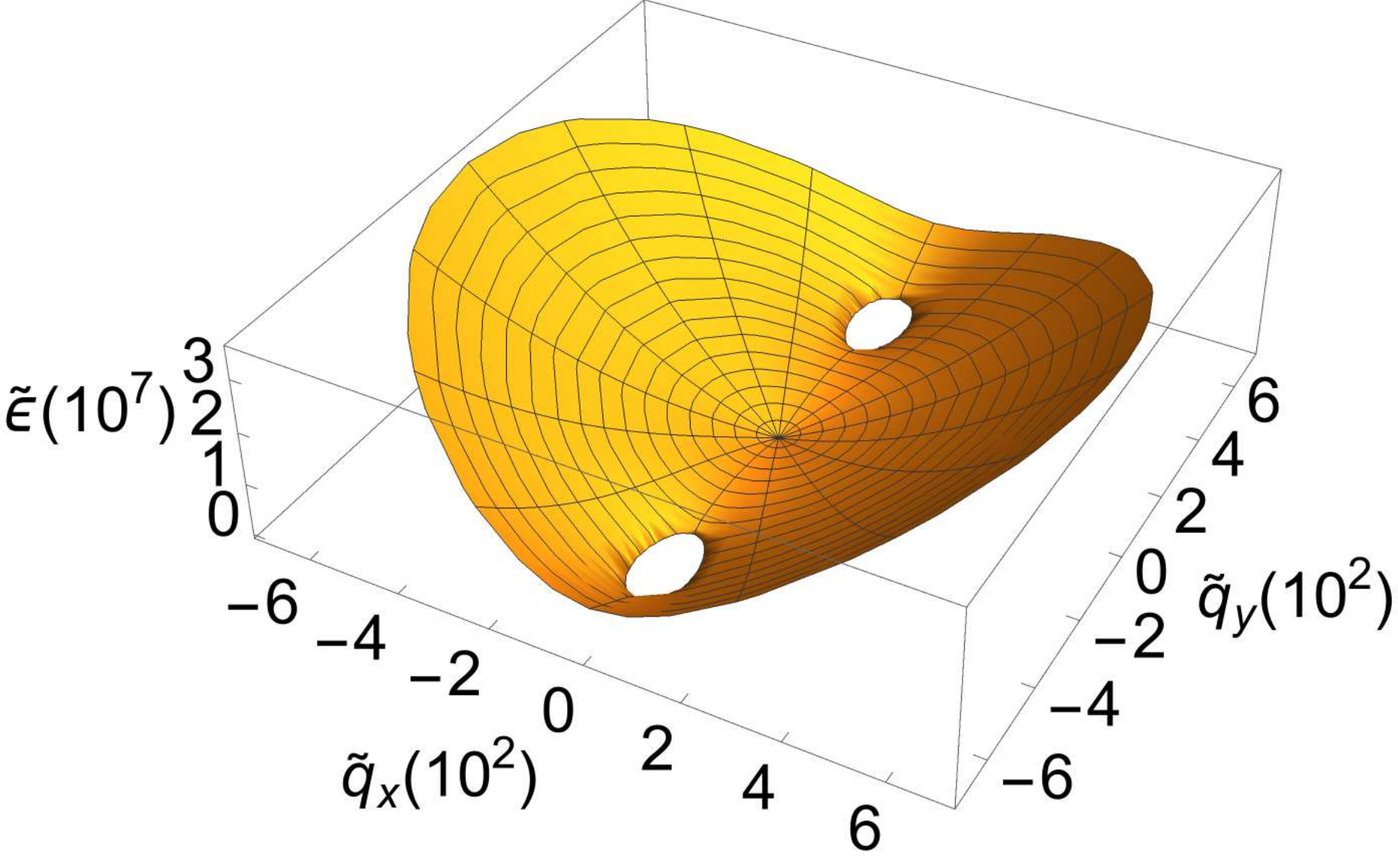}
\hspace{1.6in}
\caption{(color online) Figures showing 2D dipolar boson spectra in $q_y$ direction for tilt angles $\theta$ = 0 , 0.4, 0.9 and 1.0,
going from left to right. The first row shows 2D plots of the excitation energy squared along the $q_y$ direction for each of these values of $\theta$,
and for different scaled densities ${\tilde{n}}$ at each $\theta$. The middle row shows 3D spectra for each $\theta$ at some density
at which there is no roton instability; the bottom row shows the same but at densities at which roton instability has occurred (roton minimum is imaginary, and so shown as  holes in the 3D plot).  For $\theta$ =1 (last column), there is phonon instability at all densities}.
\end{figure*}

{\bf b.} $\theta>\cos^{-1}{\frac1{\sqrt3}}$ (= 0.955): the short range interaction becomes negative, and as a consequence, the spectrum is imaginary at small momentum. So, the system has long-wavelength ($q \rightarrow 0$) instability and hence a collapse of the BEC phase.\\

The set of 2D plots (top row) in Fig. 3 show the variation of the energy spectrum ($\epsilon^2$) vs $q_y$  with tilt angle $\theta$, and scaled density 
${\tilde{n}}$. Variation with respect  to density are shown for fixed $\theta$ = 0, 0.4, 0.9, and 1.0. The last plot is for $\theta$= 1.0, which is
beyond the critical $\theta_c$ = 0.955, and hence shows phonon instability.  In each of the other figures, with increasing density, the roton minimum
deepens, until at sufficiently large density, the frequency becomes imaginary, shown as negative $\epsilon^2$ in the plots. The middle and bottom rows
show 3D plots of the energy, $\epsilon$ vs momentum $q$ , i.e. for all directions on the plane, for two densities -- a lower density for which the roton energy is
positive (middle row) and a higher density for which the roton energy is imaginary (shown as holes on the $q=0$ plane).

\subsection{Phase Diagram}
We combine our results above for various tilt angles and densities to obtain a density vs tilt angle phase diagram for a homogeneous 2D dipolar boson system at 
zero temperature; this is shown in Fig 4. One of our goals was to see how a mean-field description compares with
Quantum Monte Carlo calculation for a homogeneous 2D Bose gas subject to DDI. Though this is at the mean-filed level, it is tempting to compare 
our phase diagram with that obtained from a QMC calculation~\cite{macia2014phase}. It is interesting that some of the key features of the QMC phase diagram are captured in our BDG calculation. While the general phase boundary of the
BEC phase, the collapse and the stripe phases are in qualitative agreement with the QMC result, we are not able to obtain the solid phase within our
BDG approximation; we find the existence of density wave/stripes in the region of the QMC solid phase of high density and smallish tilt angle.

\begin{figure}[b]
\includegraphics[width=16pc]{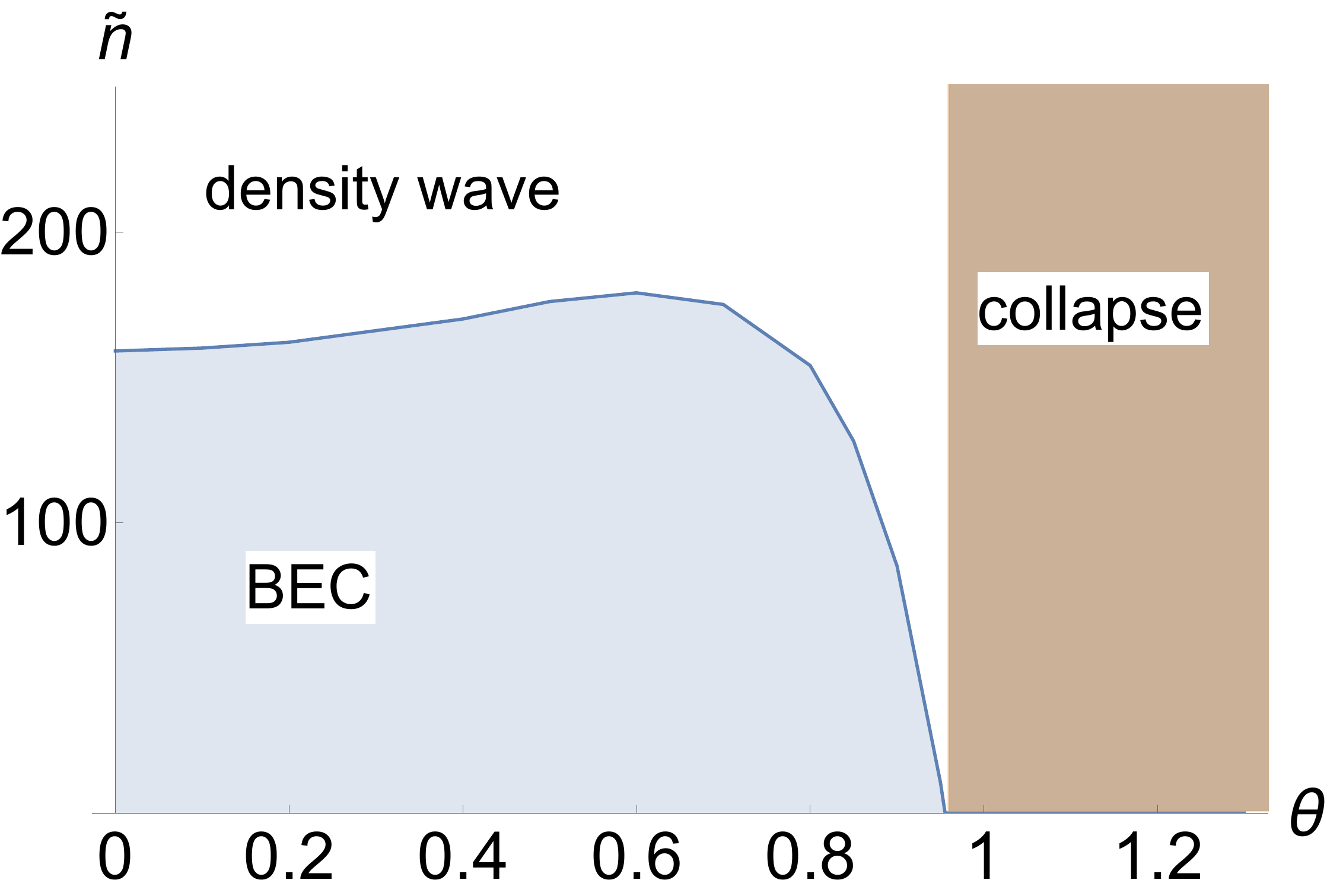}
\caption{(color online) Scaled density (${\tilde{n}}$) vs tilt angle ($\theta$)  phase diagram, based on our calculations on homogeneous 2D dipolar system.
Here, we have taken $a_{dd}$ to be $10^4a_0$, and $r_c=10a_0$.}
\end{figure}

Given that, we have attempted to 
describe the effect of a finite-range anisotropic dipole-dipole interaction in 2D, by exploring the effect of non-zero tilt angles. Our calculations and results show that
in the relatively weakly interacting regime, the roton instability does not occur, and the uniform condensate is stable. As we approach the roton instability as a function of tilt angle $\theta$ and also for large enough densities, depletion will start to increase, and at roton instability, depletion will become significant enough that the uniform condensate may not be sustainable. Since the dipolar interaction is anisotropic, it becomes attractive for some direction first, prior to becoming attractive in other directions. In our 2D case with non-zero tilt angle, this occurs first in the y-direction. Roton instability at finite momentum results in standing wave. The system exhibits density wave modulations in that direction, resulting in a "stripe phase" behavior.

\subsection{Structure factor}

We supplement our calculations of the BdG spectrum with results from our calculations of the dynamic form factor, 

The dynamic form factor is given by \cite{nozieres2018theory}
\begin{eqnarray}
	S(k,w)=\sum_n\mid< n\mid\rho_k\mid0>\mid^2\delta(\hbar w-(E_n-E_0))
\end{eqnarray}
where $\rho_k=\sum_q b^\dag_{k+q}b_q$. In BdG approximation we take only leading term, $b^\dag_qb_0+b^\dag_0b_{-q}$. Using Bogoliubov transformation,
$\rho_k=\sqrt {n_o} (u_q+v_q)(\beta^\dag_{k}+\beta_{-k})$, then
\begin{eqnarray}
	S(k,w)=\frac{n_0\hbar^2k^2/2m}{\varepsilon(k)}\delta(\hbar w-\varepsilon(k))
\end{eqnarray}
The static form factor(structure factor)
\begin{eqnarray}
	S(k)=\int dwS(k,w)=\frac {n_0\hbar^2k^2/2m}{\varepsilon(k)}
\end{eqnarray}

\begin{figure}[b]
\includegraphics[clip,width=1.5in]{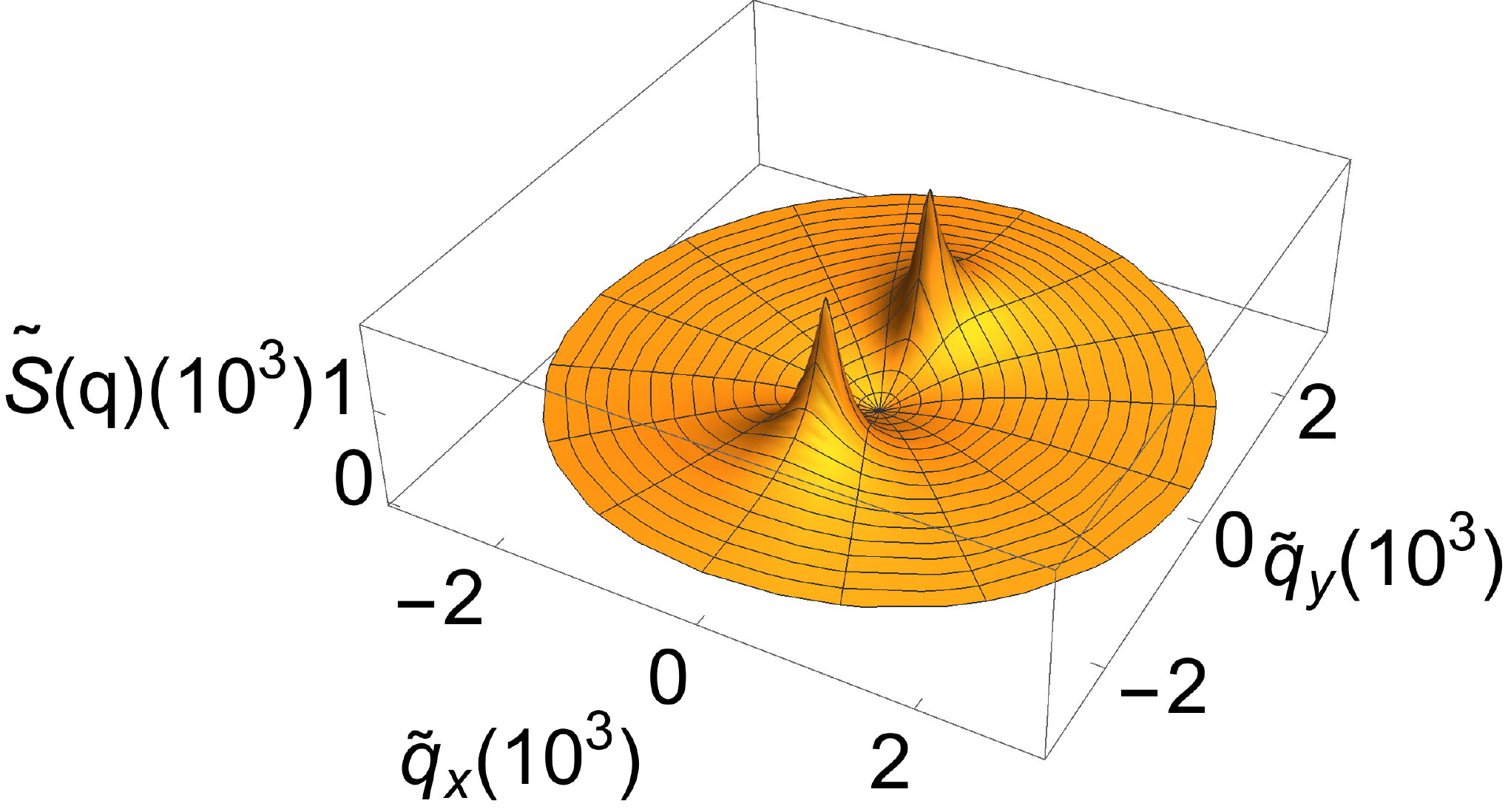}
\hspace{0.1in}
\includegraphics[clip,width=1.7in]{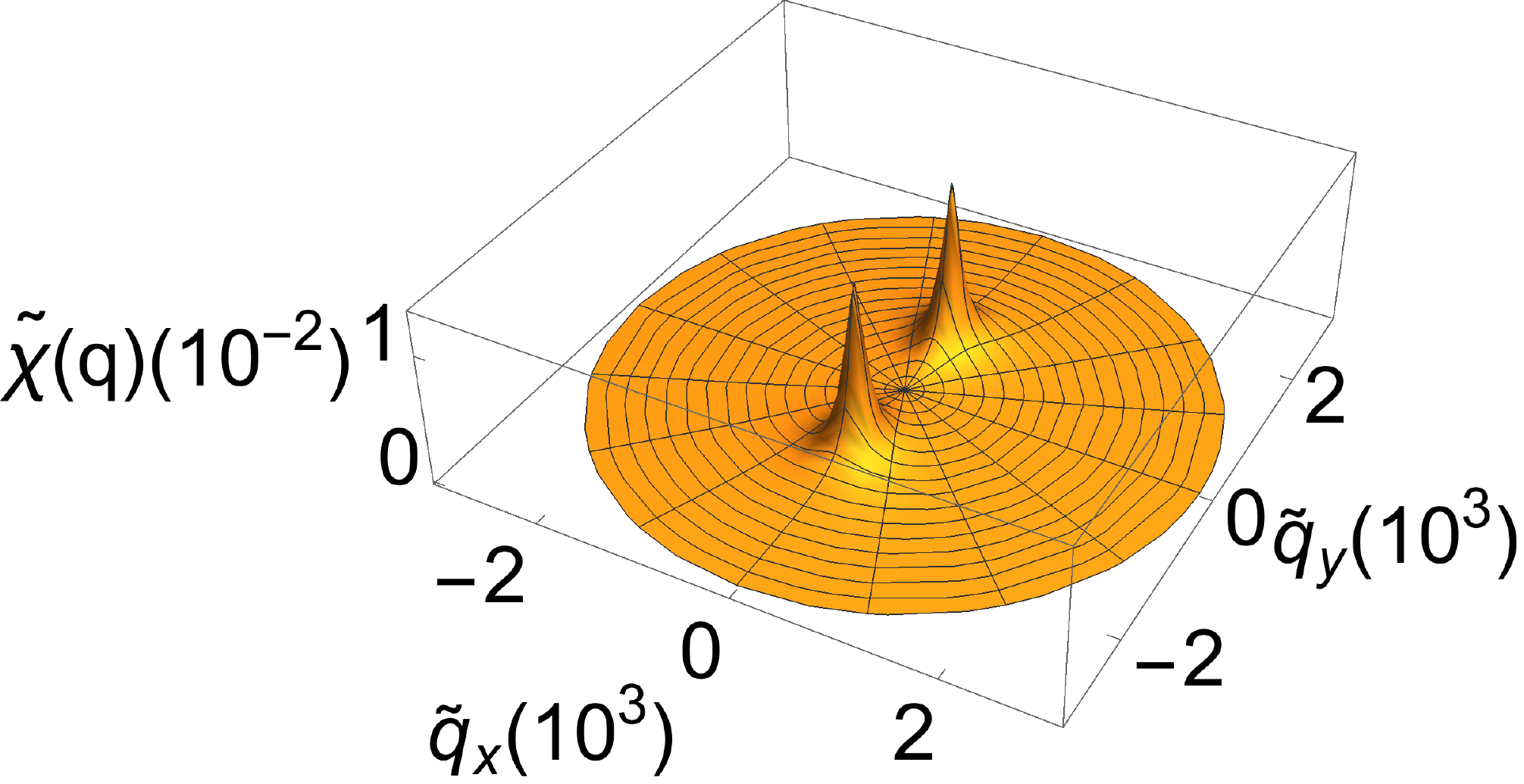}\\
\vspace{0.4in}
\includegraphics[clip,width=1.6in]{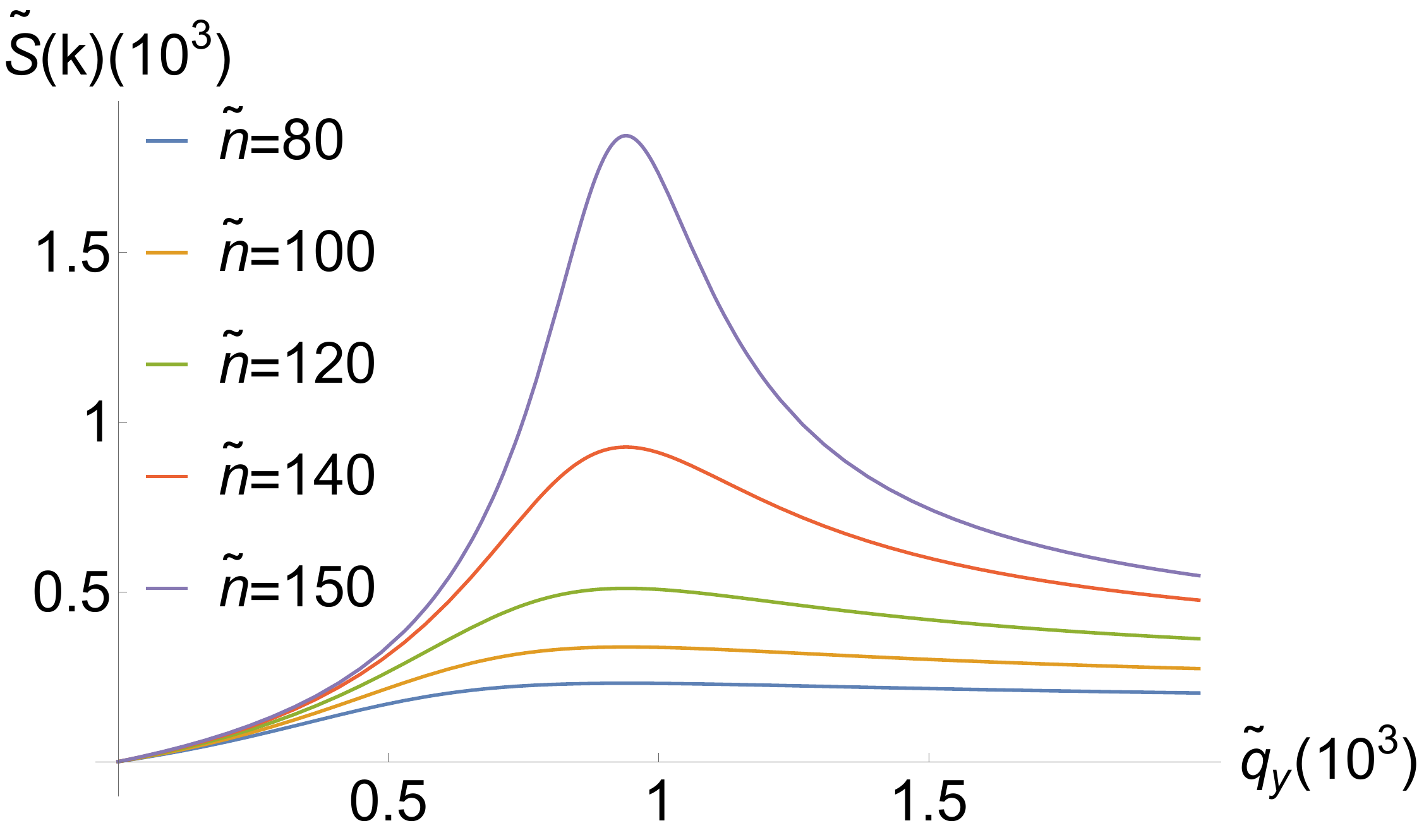}
\hspace{0.1in}
\includegraphics[clip,width=1.6in]{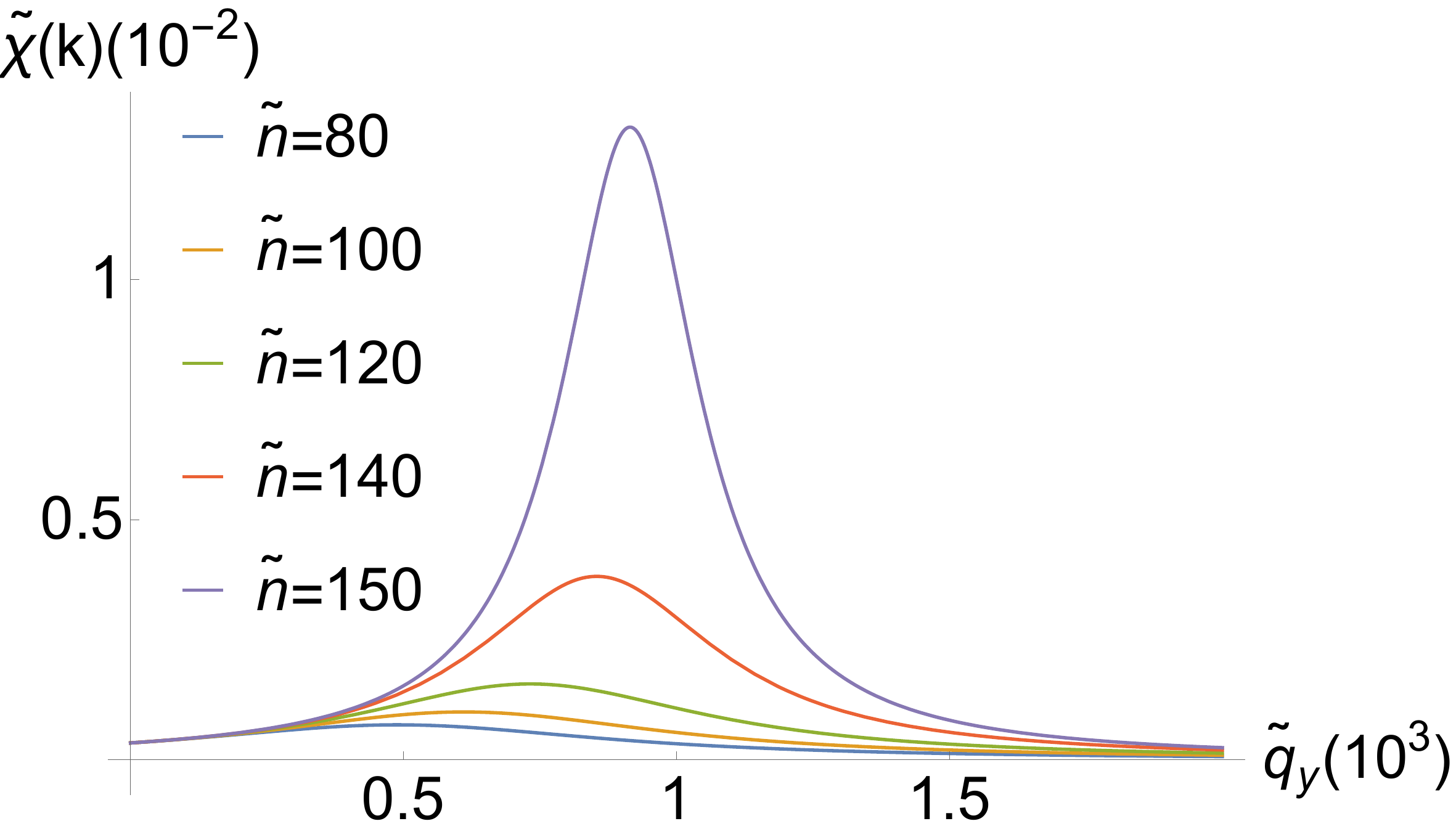}
\caption{(color online) Calculated structure factor and density-density response function for 2D dipolar boson systems. Top row figures show the 3D structure factor (left) and  density-density response function (right) at tilt angle $\theta=0.8$. The peaks are in the y-direction.
 Bottom row 2D figures show the structure factor and the density-density response in the y-direction 
 at the same tilt angle. As density increases, the response function diverges. $a_{dd}$ is taken to be $10^4a_0$  $r_c=10a_0$.}
\end{figure}

The related quantity, density-density response function is given by:
\begin{eqnarray}
\chi(k,w)=\frac{n_0\hbar^2k^2/m}{\hbar^2w^2-\varepsilon^2(k)}
\end{eqnarray}
The static density-density response function is
\begin{eqnarray}
\chi(k,w=0)=\frac{n_0\hbar^2k^2/m}{\varepsilon^2(k)}
\end{eqnarray}
The top row of Fig. 5 shows the 3D structure factor and density-density response function at a tilt angle less than the critical angle $\theta = 0.955$,
at which the collapse sets in. The peaks are in the $y$-direction. The bottom row in the figure shows the structure factor and the density-density response in 
the $y$-direction at the same tilt angle. As density increases, the response function diverges. 
Both static structure function and density-density response function diverge at {\it finite} momentum as the roton reaches zero. This is suggestive of the homogeneous BEC becoming unstable to density wave.

\subsection{Roton position and Coherence length}

From Gross-Pitaevskii equation Eq(2) without external potential, we have:
\begin{equation}
[-\frac{{{\nabla ^2}}}{{2m}}\phi_0(r) +\int dr'{\vert \phi_0(r')\vert}^2 V(r-r') ]\phi_0(r)=\mu\phi_0(r)
\end{equation}
Now, if we suppose that $\mu$ is close to the equilibrium value $nV(0)$, then we have
\begin{equation}
-\frac{{{\nabla ^2}}}{{2m}}\phi_0(r)=[n_0V(0)-\int dr'{\vert \phi_0(r')\vert}^2 V(r-r')]\phi_0(r)
\end{equation}

\begin{figure}
\includegraphics[width=70mm]{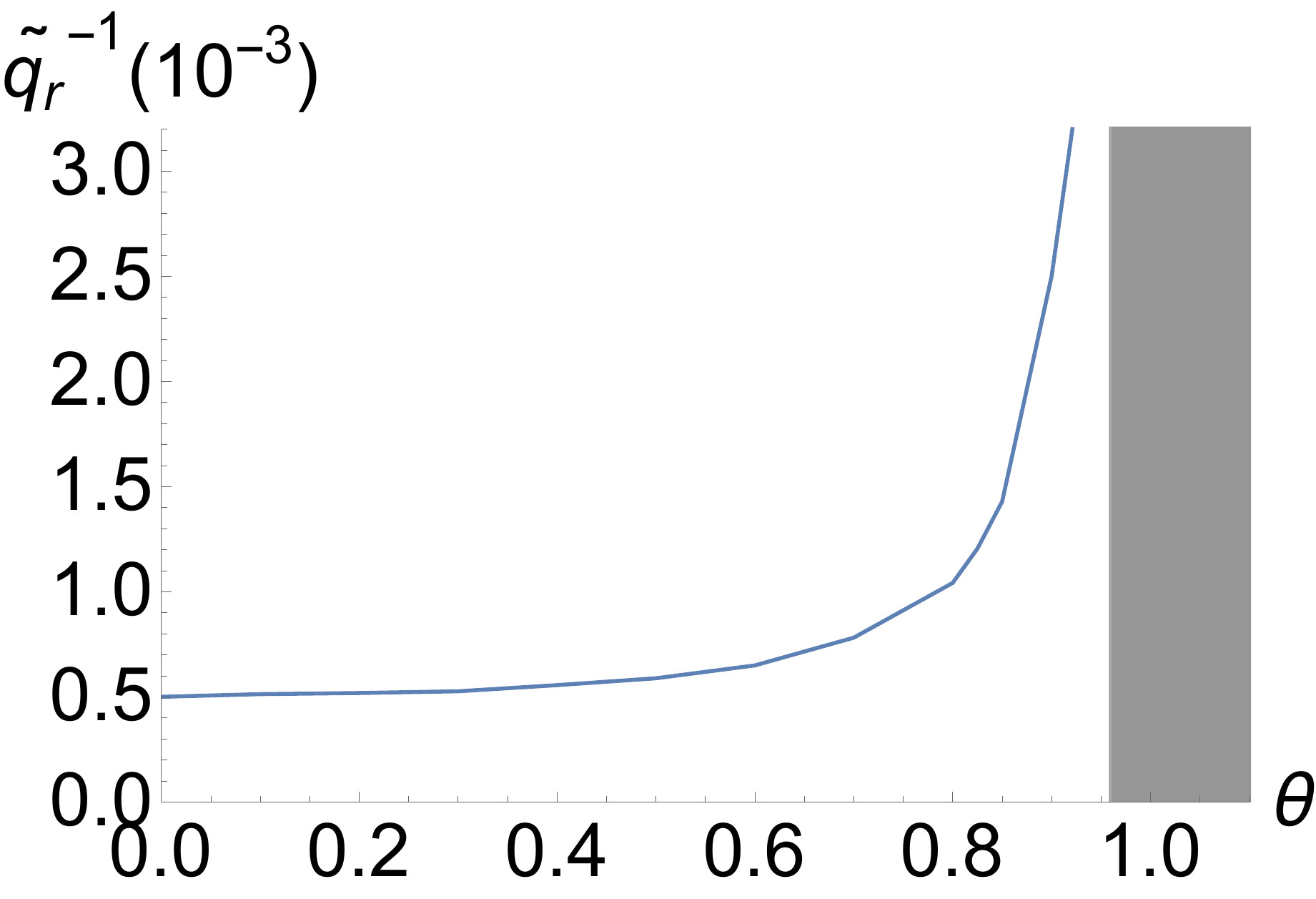}
\caption{(color online) Variation of roton position with tilt angle for 2D dipolar bosons. Figure showing the inverse of the scaled roton momentum ($1/{\tilde{q}_r}$) at roton instability, as a function of tilt angle $\theta$. As $\theta$  increases, $1/\tilde q_r$ increases.  When $\theta$ exceeds $\cos^{-1}\frac1{\sqrt3}$,  there is 
collapse to phase separation.}
\end{figure}

In this case, the spatial  variation of $\phi_0(r)$ occurs on a scale of $(mn_0V(0)-m\int dr'{\vert \phi_0(r')\vert}^2 V(r-r'))^{1/2}$. The scale over which $\phi$ varies is at least of the order $\xi=(mn_0V(0))^{-1/2}$, the coherence length.
For the 2D dipolar interaction, $\xi=(mn_0V(0))^{-1/2}=2\pi\frac{ mn_0d^2}{r_c}{P_2(\cos\theta)}$. As $\theta$ increases, the coherence length $\xi$ decreases, and as a result, the roton instability also occurs at a smaller momentum. This result is summarized in Fig. 6.

\begin{figure*}
\includegraphics[clip,width=1.9in]{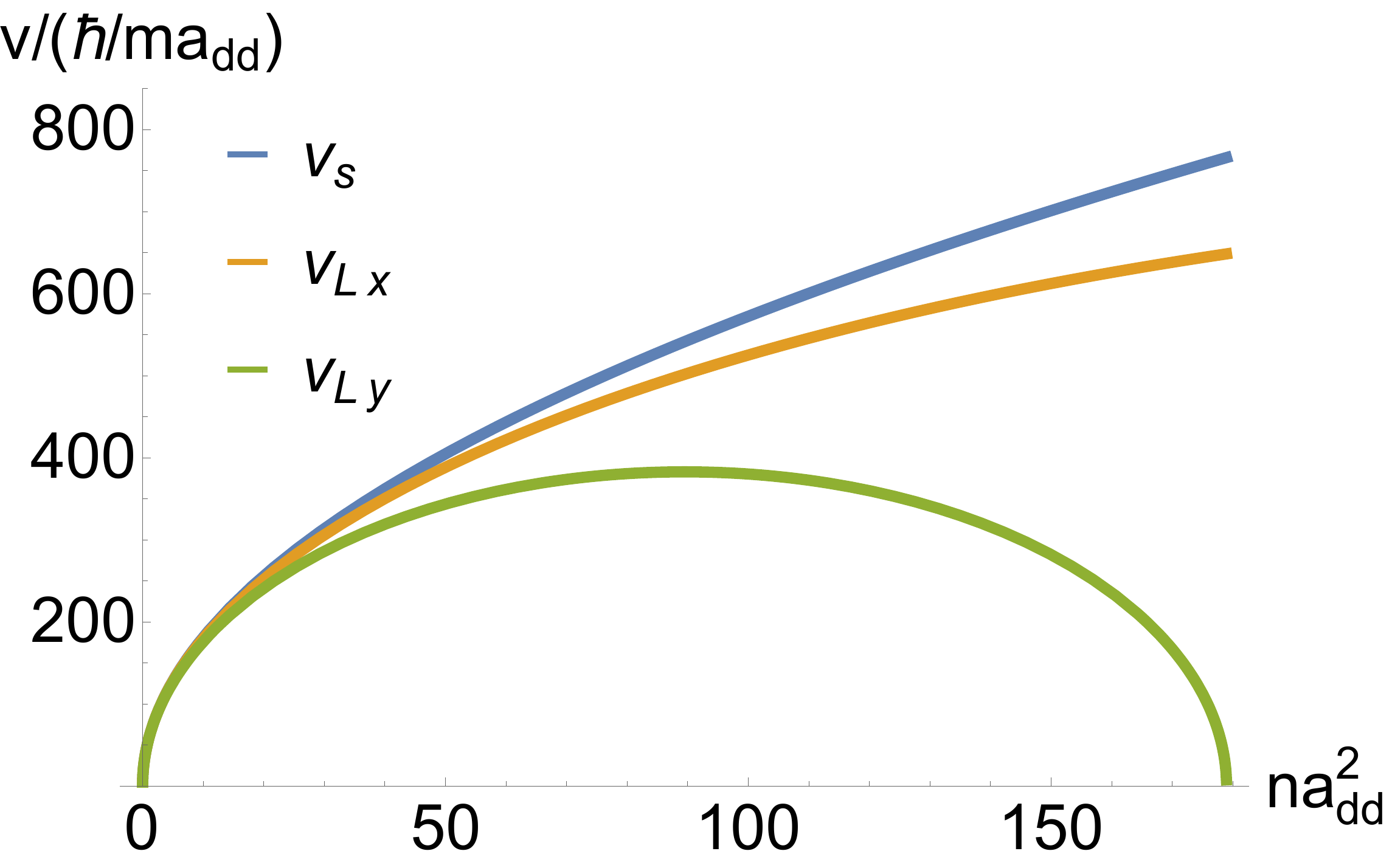}
\hspace{0.2in}
\includegraphics[clip,width=1.9in]{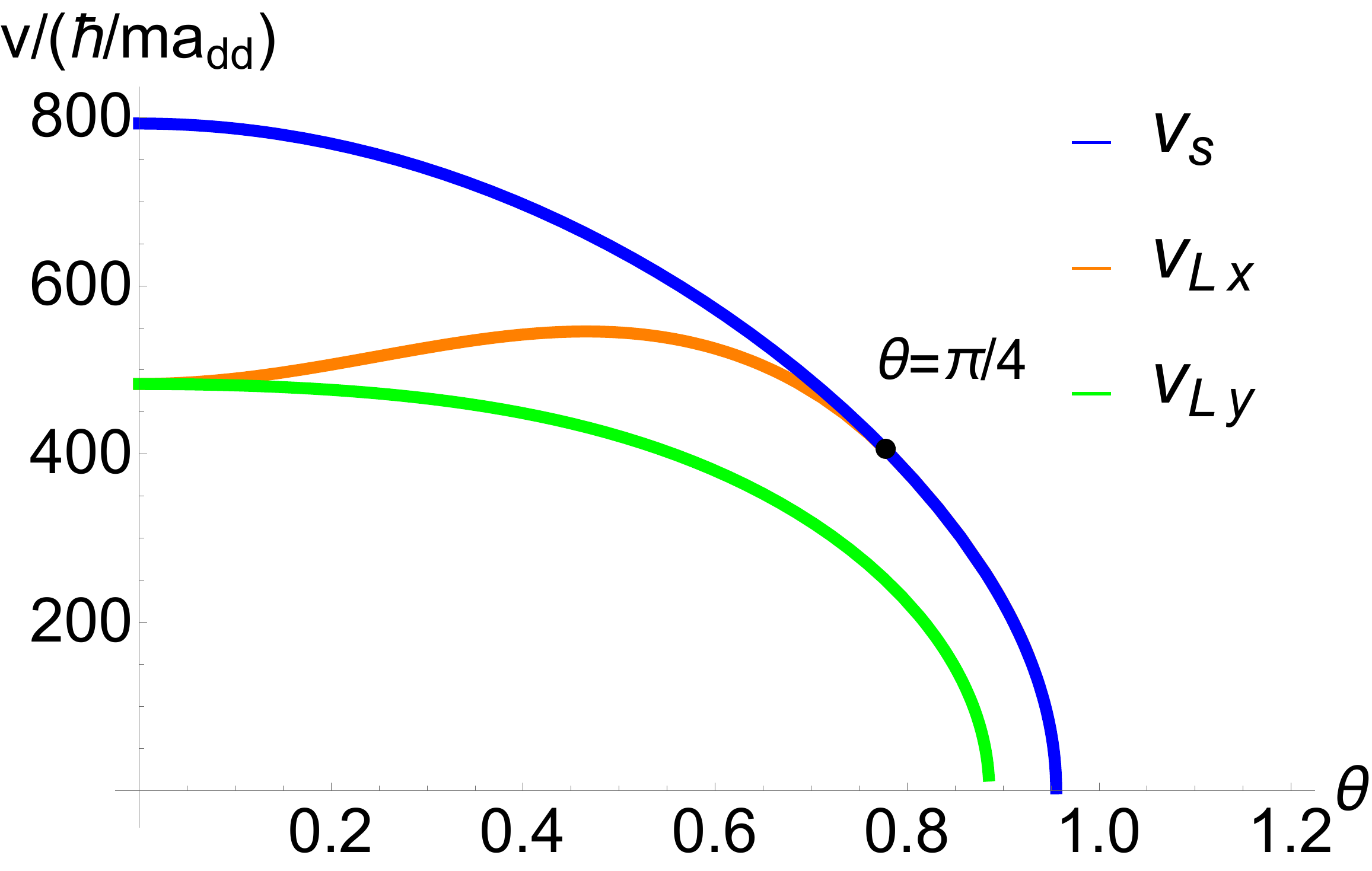}
\hspace{0.2in}
\includegraphics[clip,width=1.9in]{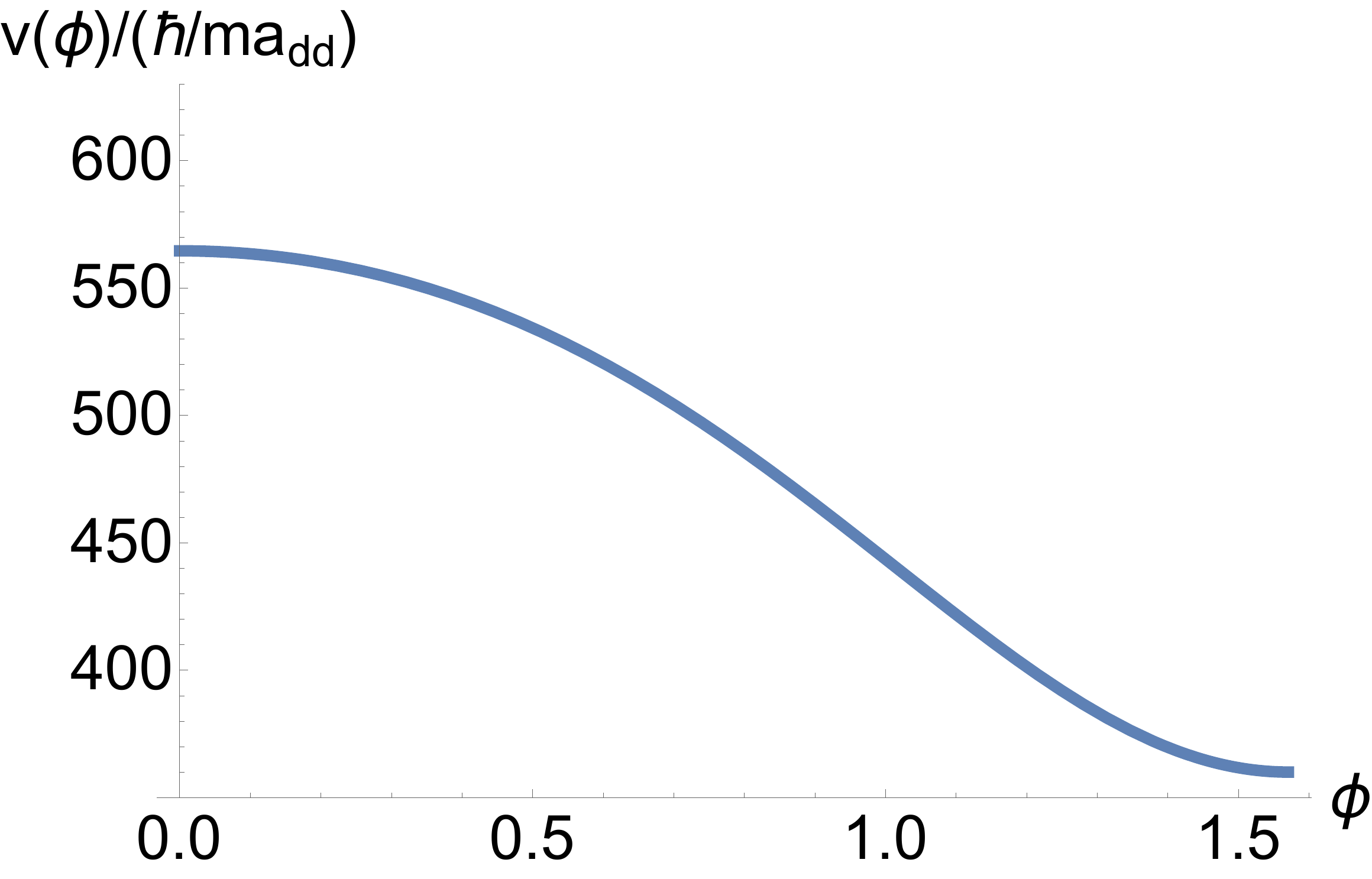}
\caption{(color online) Calculated sound and Landau critical velocities in homogeneous 2D dipolar boson system. Left plot shows Landau velocity $v_L$ in x and y direction, as well as sound velocity, $v_{sound}$ as a function of density at fixed $\theta=0.6$ . 
The  middle plot shows $v_L$ and $v_{sound}$ as tilt angle $\theta$ is varied. varying. Starting at $\theta=\pi/4$,  $v_{Lx}=v_{sound}$.
The right plot shows the variation of $v_L$ with in-plane angle $\phi$ , i.e. from x direction ($\phi=0$) to y direction ($\phi=\pi/2$).} 
\end{figure*}

\subsection{Sound velocity and Landau critical velocity}
The sound velocity of boson gas is given by, $v_{sound}=\lim_{k\to0}\frac{\varepsilon(k)}{\hbar k}$, It is isotropic for dipolar boson gas even though the interaction is anisotropic, since the short range part of the interaction is still isotropic. Thus, $v_{sound}=\hbar^2/(ma_{dd})\sqrt{(3\cos ^2\theta-1)\pi \tilde n a_{dd}/{r_c}}$. As tilt angle $\theta$ increases, the short range interaction decreases, and so does $v_{sound}$. At $\theta=\cos^{-1}\frac{1}{\sqrt3}$, $v_{sound}$ becomes zero. When $\theta>\cos^{-1}\frac{1}{\sqrt3}$, the system collapses, i.e. undergoes a phonon instability, and $v_{sound}$  then becomes imaginary.

The critical velocity in a superfluid is given by the Landau criteria, $v_L= {\text {min}} (\frac{\varepsilon(k)}{\hbar k})$.~\cite{landau-sf} For the spectrum of BEC gas without roton-maxon feature, $v_L=v_{sound}$. The existence of roton decreases $v_L$; $v_L={\frac{\varepsilon(k)}{\hbar k}}\vert_{k=k_{roton}}$,  where $k_{roton}$ is the minimum of the roton in the spectrum. When roton reaches zero, $v_L=0$, and there is no superfluity. The Landau critical velocity, 
$v_L$ is anisotropic: In the $y$-direction, where the interaction is negative, roton lies lower; $v_L$ is smaller than in other directions.  The dipolar interaction, $V_l(q)$ is always negative in the $y$-direction.  In the $x$-direction and near it, when $\theta$ is larger than some value, $V_l(q)$ becomes positive, and consequently there is no roton; $v_L= v_{sound}$ after that certain angle. Our results for the homogeneous 2D dipolar bosons are shown in Fig. 7.





\section{Results: Quasi-2D Dipolar Bosons}

\begin{figure*}
\vspace{0.4in}
\includegraphics[clip,width=1.6in]{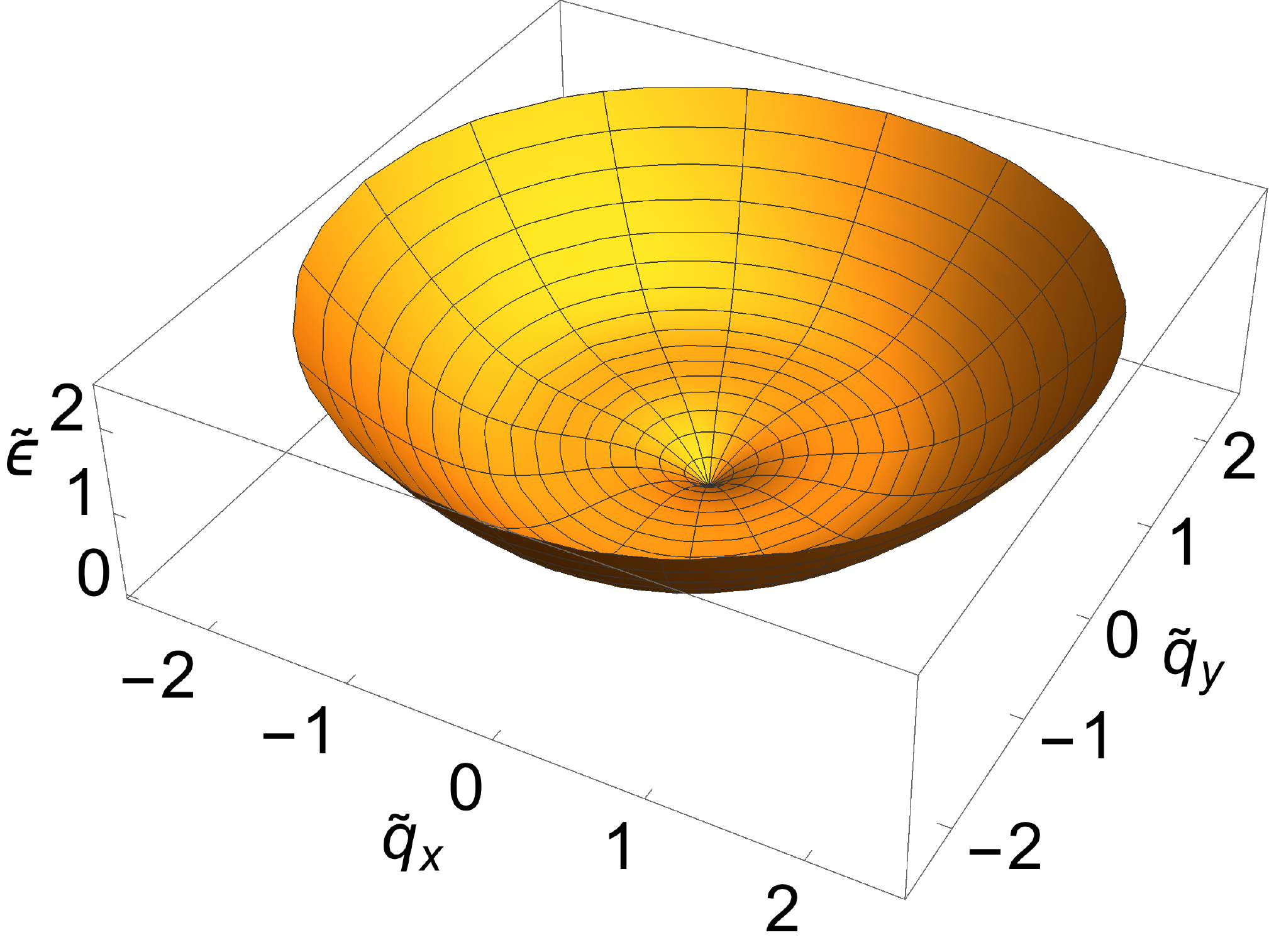}
\includegraphics[clip,width=1.5in]{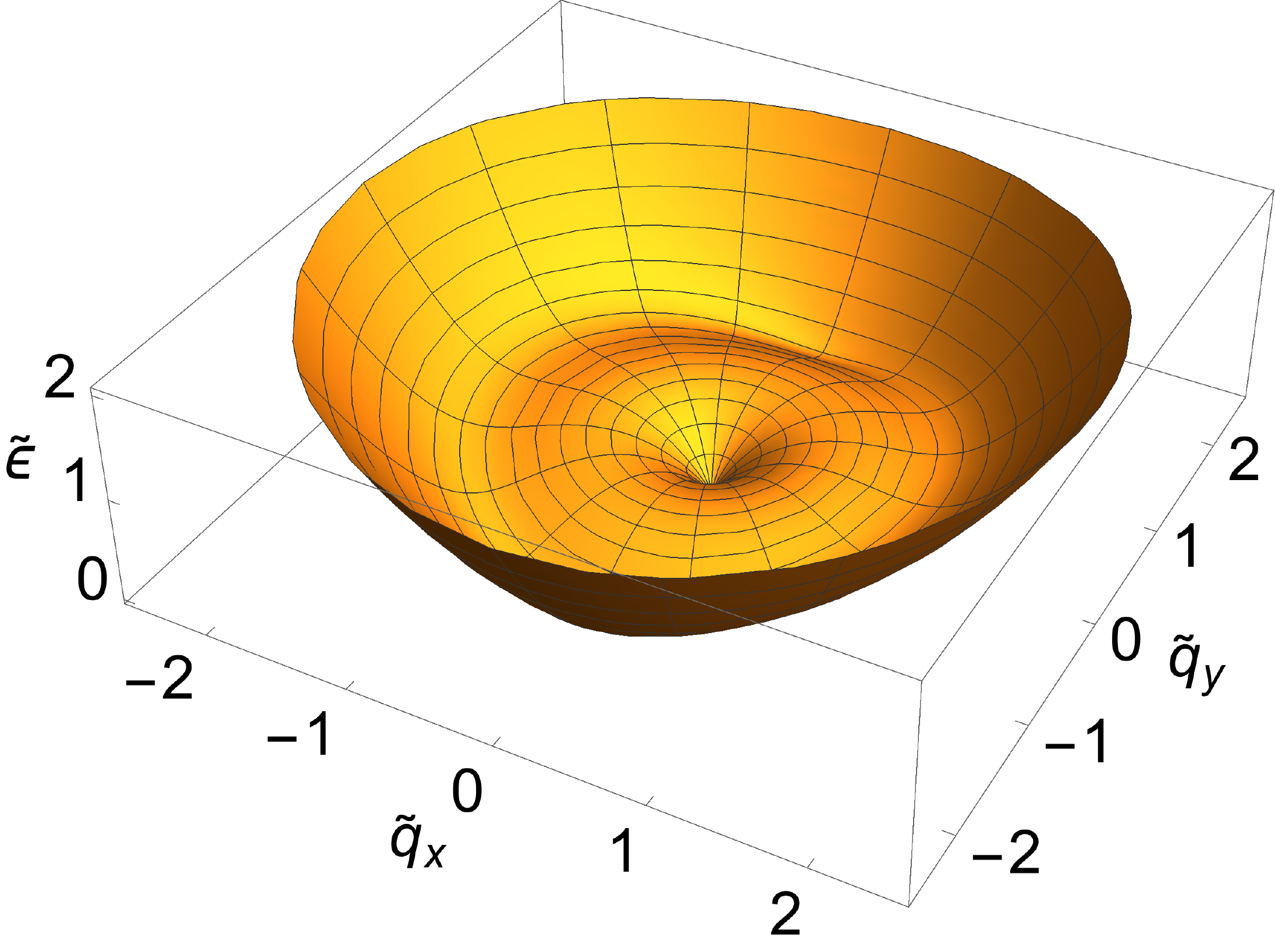}
\includegraphics[clip,width=1.5in]{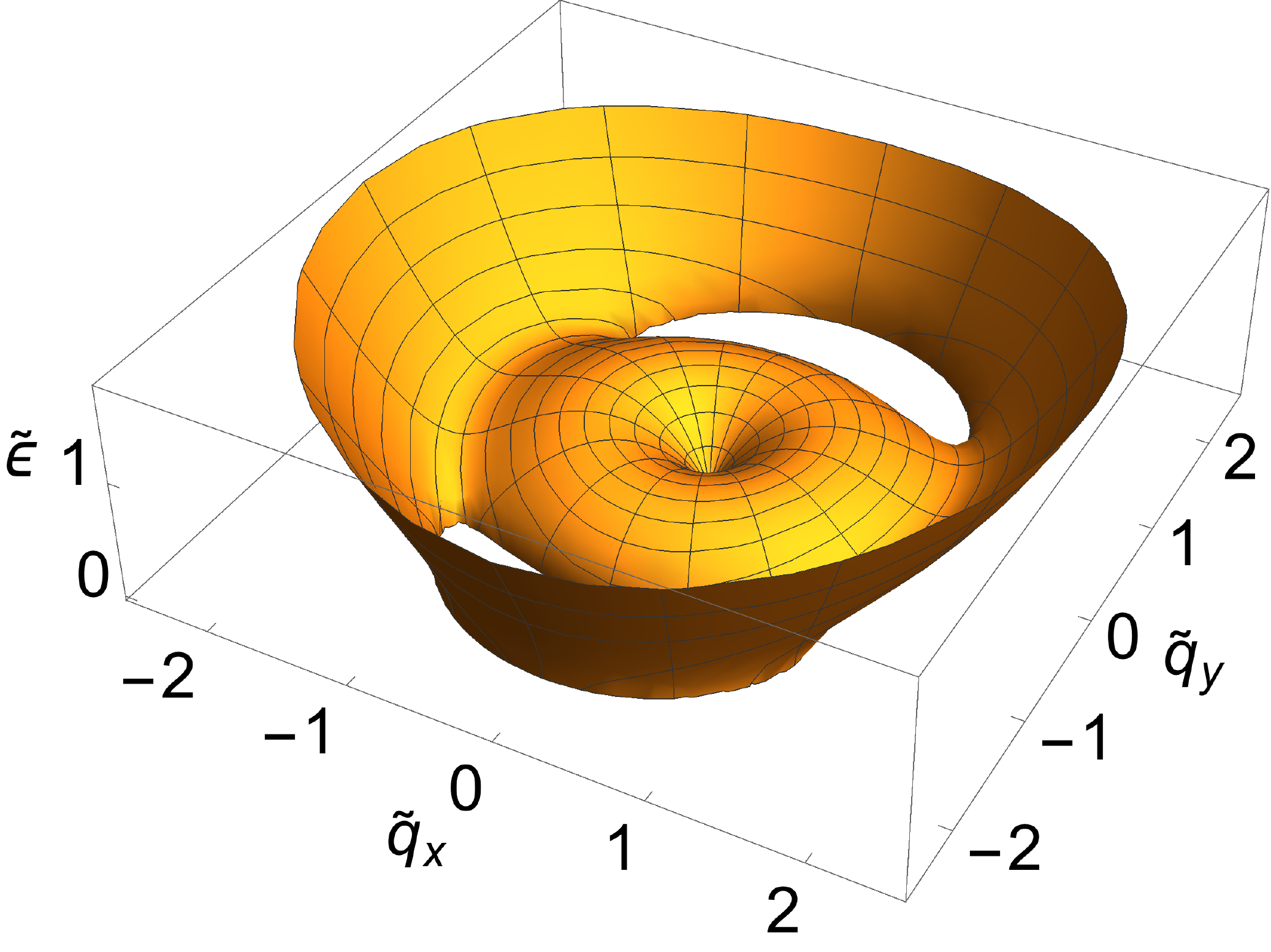}
\includegraphics[clip,width=1.5in]{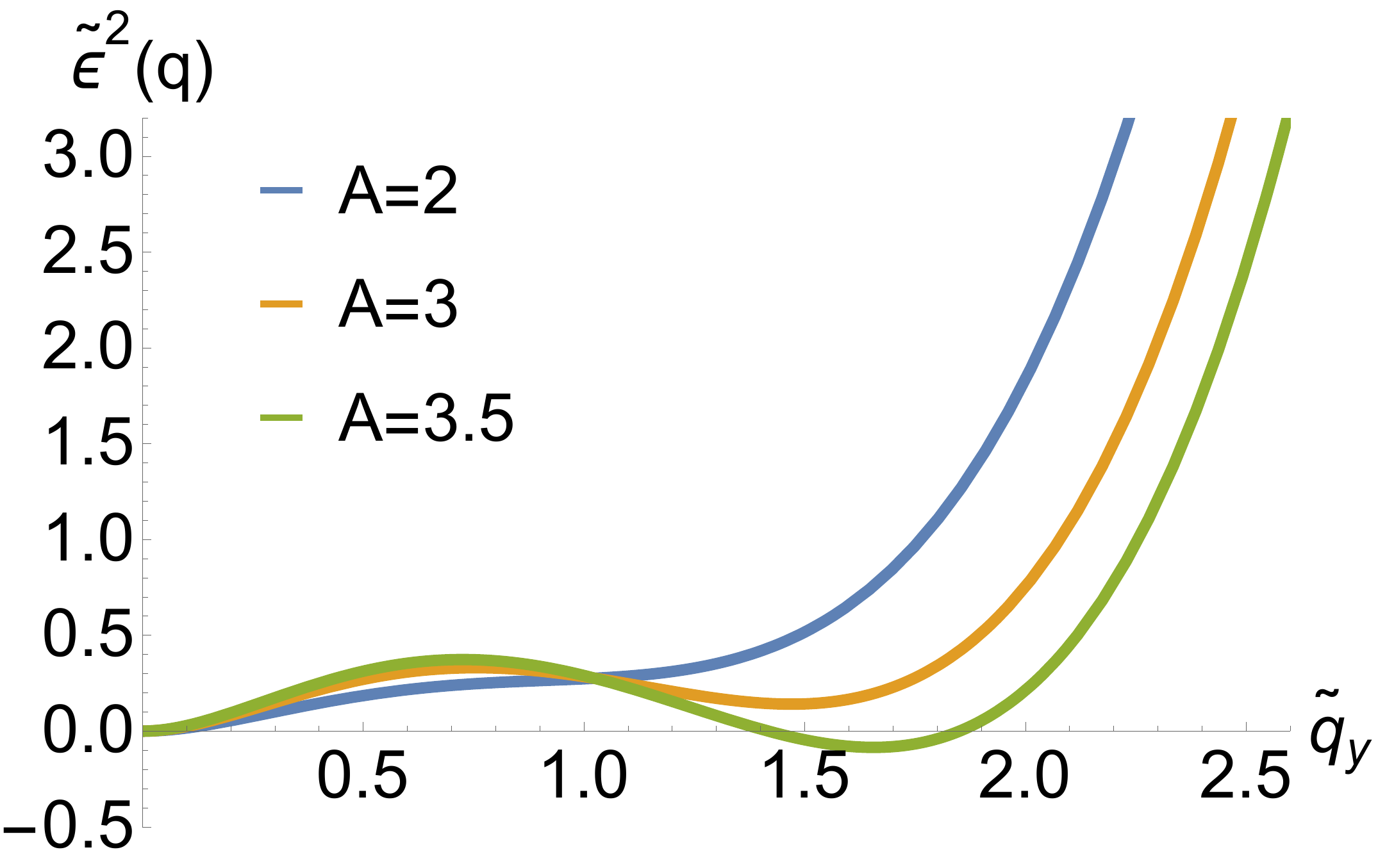}
\caption{(color online) Calculated excitation spectra of our quasi-2D dipolar boson system. From left to right, the figure shows the 
3D spectrum as the parameter A  (defined in text) is increased, i.e.  A= 2,3,3.5 respectively. for a fixed tilt angle $\theta=0.2$, 
The  right--most plot shows the respective spectra in the $y$-direction }
\end{figure*}

\subsection{Spectrum and Instability}

Using Eq.(17), the Quasi-2D BDG spectrum is given by $\varepsilon(k)= [(\hbar k^2/{2m})^2+n_0V_{2D}(k) \hbar k^2/{2m}]^{1/2}$.
We scale the spectrum with the trap frequency $w_z$, so $\epsilon(k) \rightarrow  \epsilon(k)/\hbar w_z$, and scale momenta so that $k \rightarrow kd_z$,
where $d_z =\sqrt{\hbar/mw_z}$.
Defining a  parameter $A = \mu_{2D}/\hbar w_z$, we note that the 2D density is related  to $A$ by $A=n_0V_{2D}(0)/\hbar w_z$.
Thus, the spectrum attains a form containing the tilt angle $\theta$, and azimuthal angle $\phi$,

\begin{widetext}
	\begin{eqnarray}
	\epsilon (k, \theta)=\sqrt{\frac{k^4}{4}+A k^2 \left(1-\frac{2 F(k) \left[\cos ^2(\theta )-\sin ^2(\theta ) \cos ^2(\phi )\right]}{\frac{8 P_2(\cos\theta)}{\left(3 \sqrt{2 \pi }\right) }}\right)}
	\end{eqnarray}
\end{widetext}

The quasi-2D condition, $\mu_{2D}\ll\hbar\omega_z$, i.e. $A\ll 1$ imposes restriction on the range of density in quasi-2D experiments. If density is too large, strong interaction will excite the particle out of the ground state of the trap, the system will be in the quasi-2D to 3D crossover regime.

As in the homogeneous 2D case, for $\theta<\cos^{-1}(1/\sqrt3)$,  the spectrum is stable, and as density increases, a roton minimum develops in the $y$-direction. At sufficient large density, the spectrum becomes unstable as the roton minimum touches zero, and the spectrum becomes imaginary.
When $\theta<\cos^{-1}(1/\sqrt3)$, the interaction is attractive, and spectrum is imaginary at any density; the system develops a phonon instability.

It may be interesting to consider if the stripe phase would occur in quasi-2D dipole system, as it does in the homogeneous 2D case.
To understand this, we follow Fischer~\cite{Fischer} and examine the quasi-2D condition, $\mu_{2D}\ll\hbar\omega_z$, i.e. $A\ll 1$, but here 
as a function of the tilt angle $\theta$. At $\theta=0$, the critical value of $A$ was found~\cite{Fischer}  to be $A_c=3.446$.
This indicated that a quasi-2D purely dipolar system of bosons is always stable, unless the density is sufficient for the systems to
be in quasi-2D to 3D crossover regime. We determine the critical $A_c$  as a function of $\theta$ from the value of A at which the roton minimum reaches
zero (see Fig. 8). In Fig. 9, we plot $A_c$ vs tilt angle $\theta$. This shows that the quasi-2D dipole gas is a stable BEC over a range of $\theta$, because $A_c$ is outside quasi-2D region. At large title angle close to $\cos^{-1}(1/\sqrt3)$, $A_c$ is small enough and inside the  quasi-2D region, which means it is possible for a quasi-2D dipole system with that title angle to be unstable towards a density wave phase. At title angle larger than  $\cos^{-1}(1/\sqrt3)$, the system is unstable toward collapse, i.e. phonon instability.

\begin{figure}[b]
\includegraphics[width=70mm]{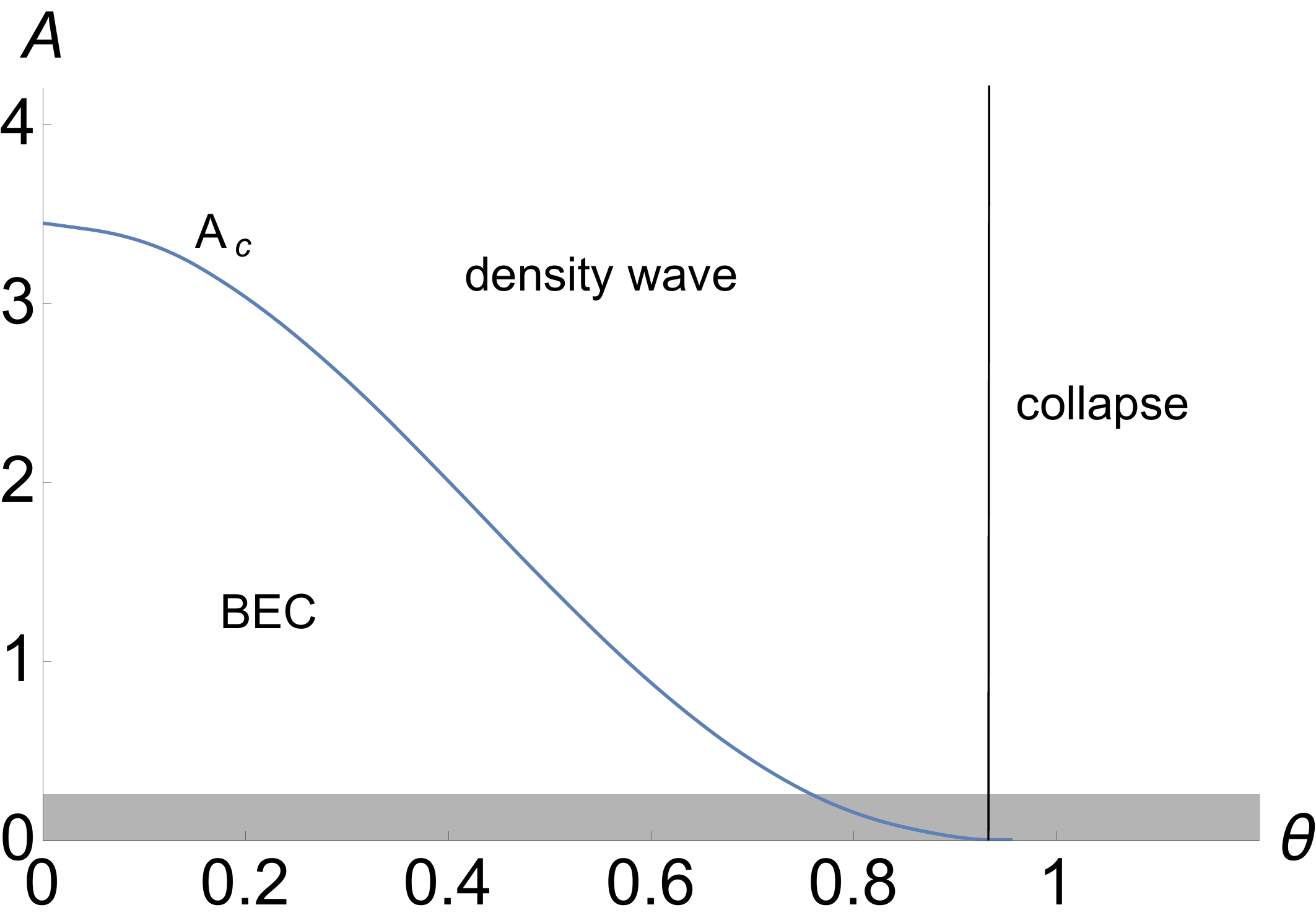}
\caption{(color online) Calculated critical value ($A_c$) of the quasi-2D parameter A  plotted as a function of tilt angle $\theta$. 
The shaded part is meant to schematically show the strictly quasi-2D region in this A vs $\theta$ phase diagram. The
line of collapse instability is also shown.}
\end{figure}

\subsection{Structure factor}

\begin{figure}
\includegraphics[clip,width=1.5in]{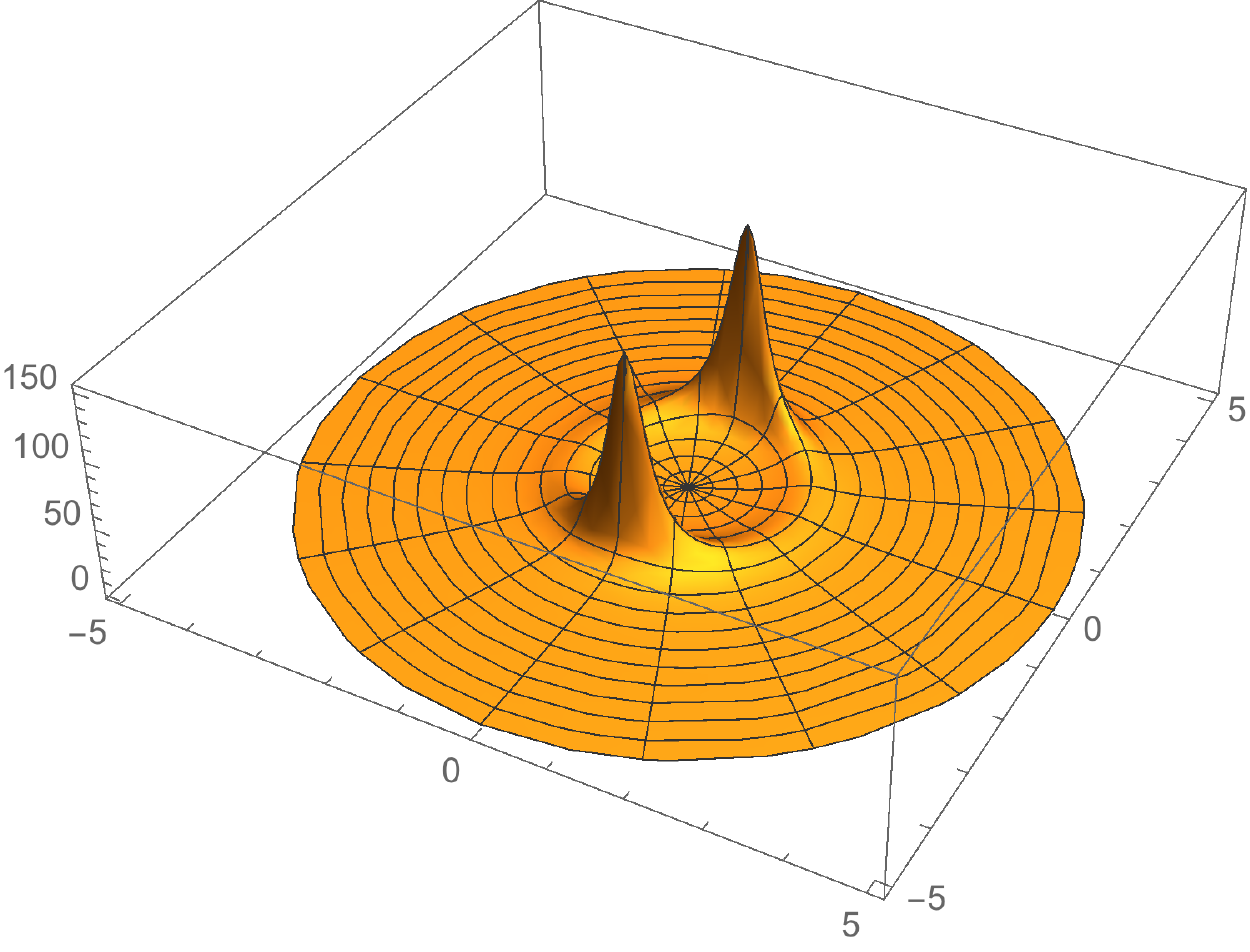}
\hspace{0.1in}
\includegraphics[clip,width=1.7in]{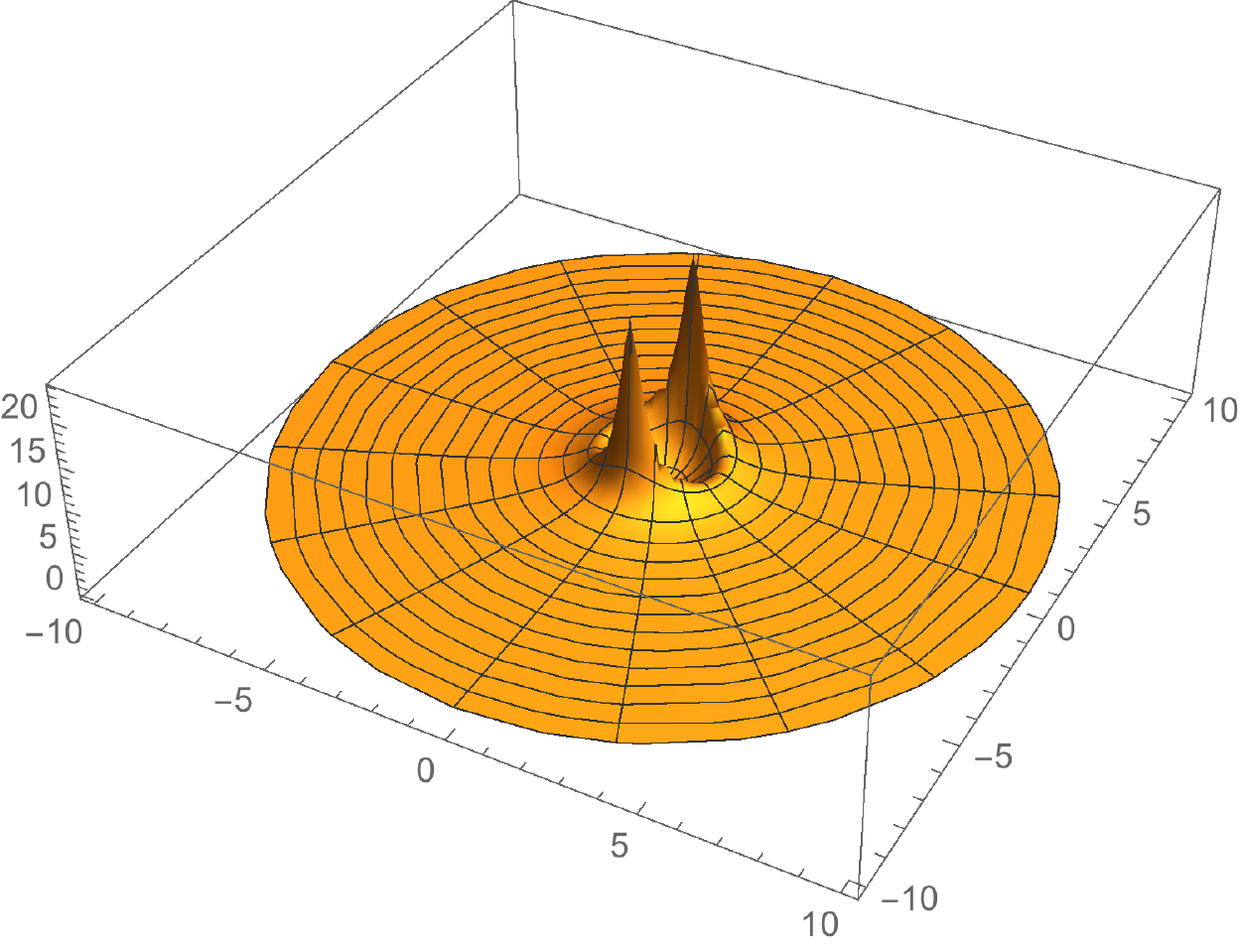}\\
\vspace{0.2in}
\includegraphics[clip,width=1.6in]{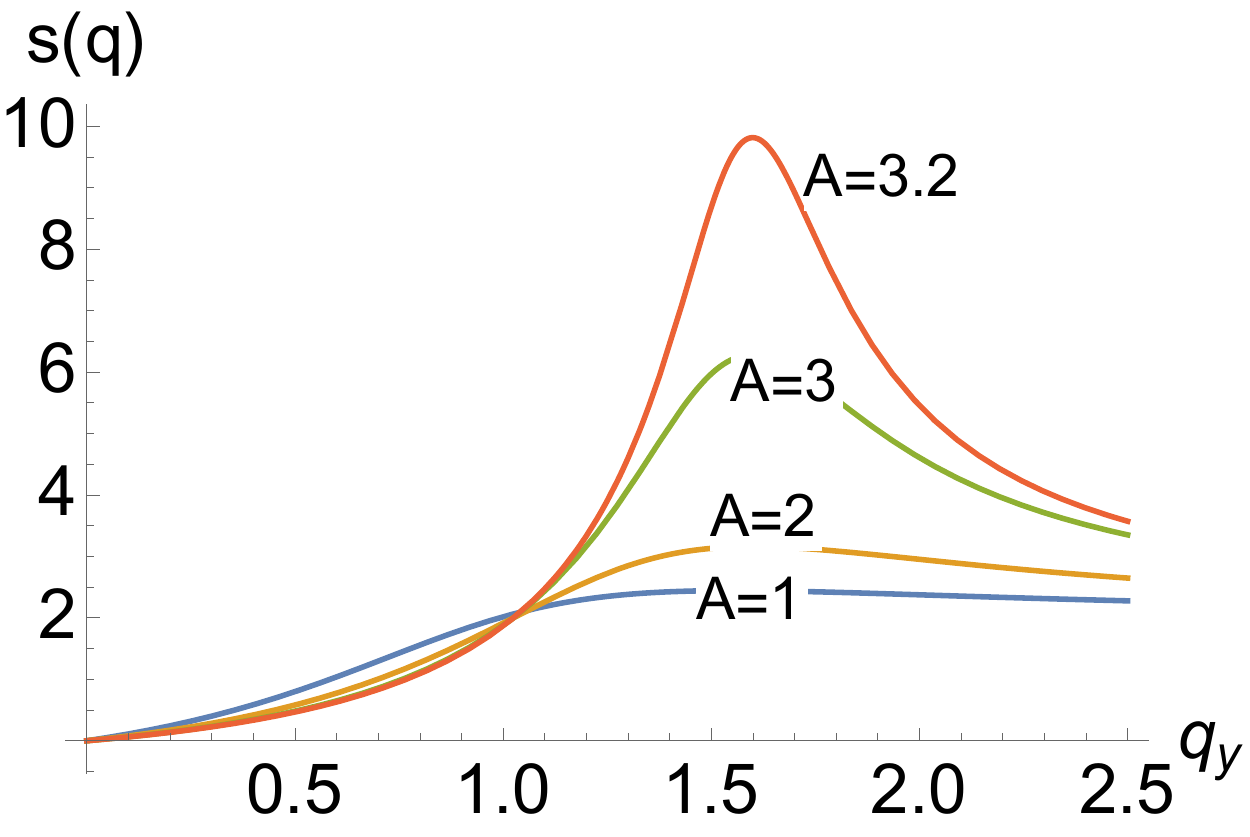}
\hspace{0.1in}
\includegraphics[clip,width=1.6in]{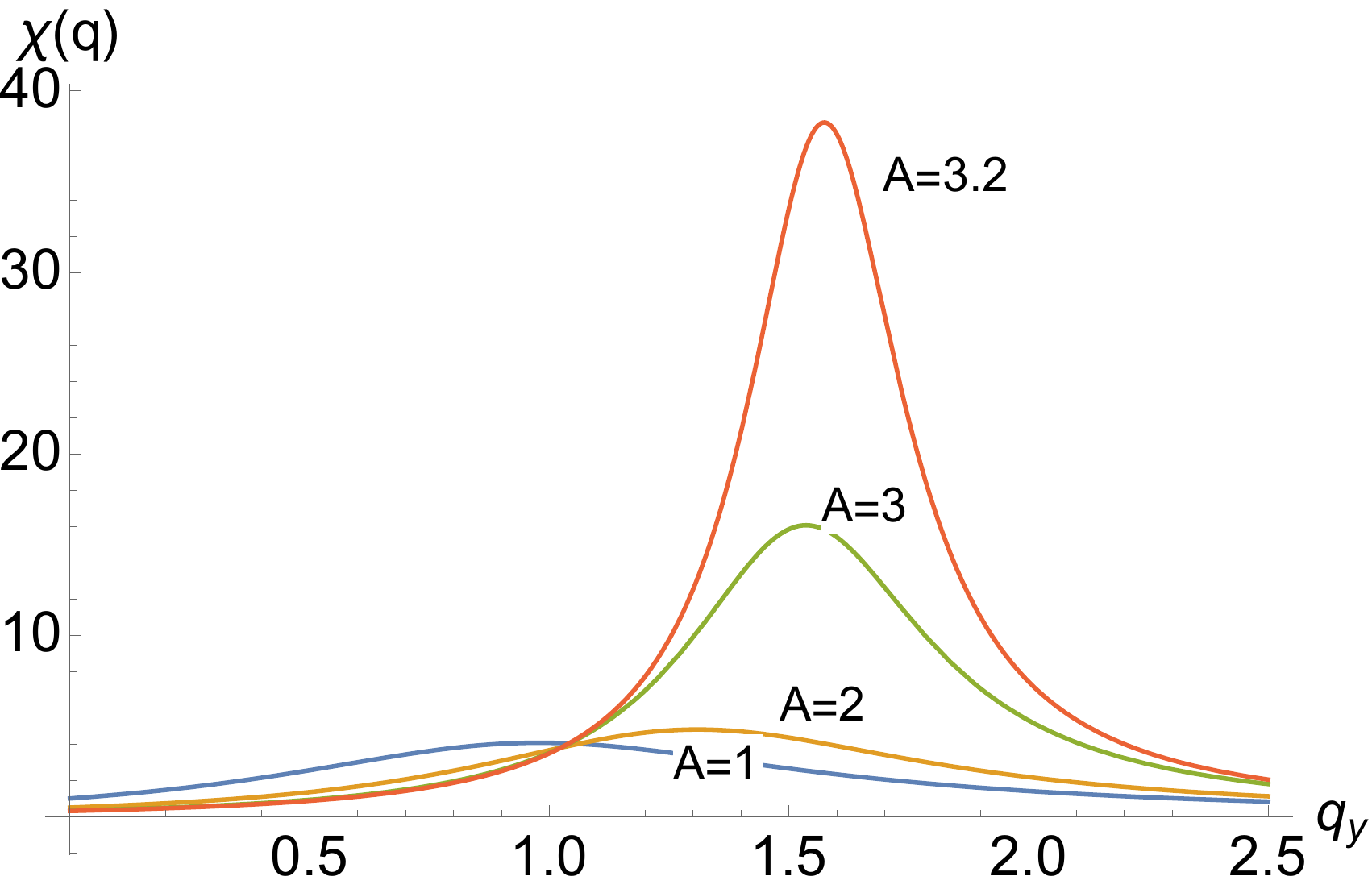}
\caption{(color online) Calculated structure factor (left) and density-density correction function (right) in quasi-2D dipolar boson system at $\theta=0.2$. The bottom figures shows the variation of these quantities with the quasi-2D parameter A. These diverges as A approaches the critical value, $A_c$.}
\end{figure}

The structure factor and density-density correction function in quasi-2d dipole boson are obtained in a similar manner to the 2D case. 
The figures below show both the 3D plots as well as the  variation of these quantities with A.

\subsection {Quantum depletion}

It is well established that at T=0, Bose-Einstein condensation occurs in 2D, with or without a trap. There can however be excitations
out of the condensate owing to quantum effects; this is referred to as {\it quantum depletion}. At non-zero T, there are additional thermal excitations.
The excited boson occupation number $n_p$ = $<b_p^\dag b_p>$ can be obtained from the Bogoliubov quasiparticle operators $\beta_p$, $\beta_p^{\dag}$, vis  Bogoliubov transformation. Thus at finite-T, 

\begin{eqnarray}
n_p=<b^\dag_pb_p>=v_p^2+(u_p^2+v_p^2)<\beta^\dag_p \beta_p> .
\end{eqnarray}
$<\beta^\dag_p \beta_p>$ follows Bose distribution, i,e.  $<\beta^\dag_p \beta_p>=\frac 1{e^{\beta \varepsilon_{p}}-1}$
At zero temperature, only first term still contributes, giving quantum depletion; the second term only exists at non zero temperature and is call thermal depletion.

\begin{figure}
\includegraphics[width=70mm]{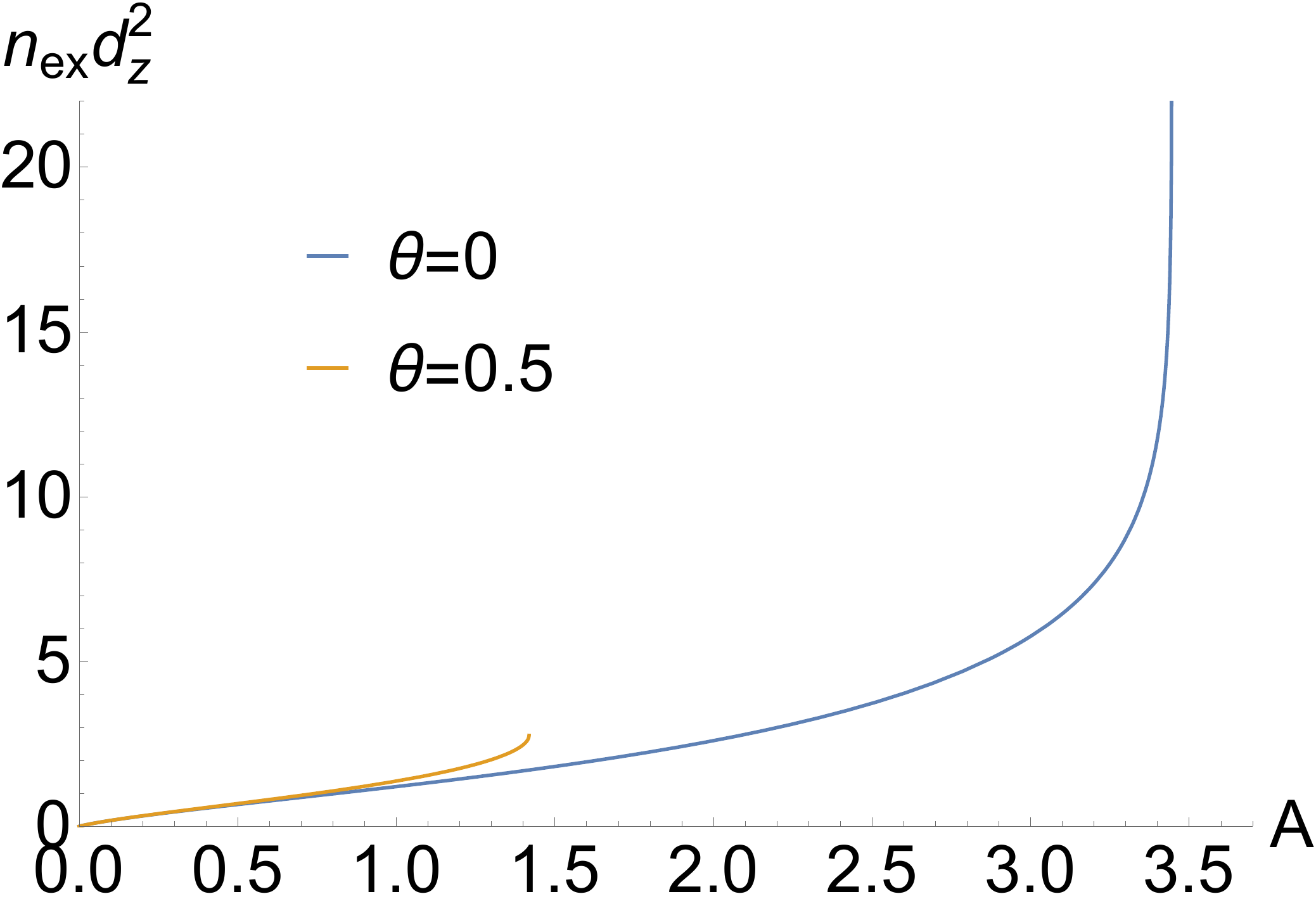}
\caption {(color online) Calculated quantum depletion for our quasi-2D dipolar boson system, as a function of the quasi-2D parameter A, for tilt angles $\theta=0$ and $\theta=0.5$. The depletion increases as A approaches a critical value. At $\theta=0$, the depletion diverge at $A_c$; for $\theta=0.5$, the depletion is finite at $A_c$.}
\end{figure}

Thus, the total quantum depletion is given by
\begin{eqnarray}
n_{ex}= n - n_o= \int \frac {d^dp}{(2\pi)^d}\frac{\epsilon(q)}{\varepsilon(q)}
\end{eqnarray}
where, $\varepsilon(q)$ is BDG spectrum and $\epsilon(q)=\hbar^2q^2/2m+n_oV(q)$ is the difference between the Hartree-Fock energy of a particle and the chemical potential. The depletion diverges at roton minimum momentum when the minimum reaches zero. 

We calculate the quantum depletion of quasi-2D dipolar bosons.
For zero tilt angle, $\theta=0$, the spectrum is isotropic, the roton minimum is a ring at $\textbf q=q\hat r$. The total quantum depletion diverges when the roton minimum reaches zero. That shows that the BDG treatment becomes invalid near the transition.
For non-zero tilt angle, $\theta \neq0$, the spectrum is anisotropic, The roton minimum occurs two points at $\textbf q=\pm q\hat y$. The total quantum depletion remain finite when the roton minimum reaches zero. The BDG treatment is valid even near the transition density. Fig. 11 shows our calculated 
quantum depletion for $\theta =0$ and $\theta=0.5$.

\section{Discussion}

In this paper we have attempted to describe a number of  zero temperature properties of homogenous 2D and quasi-2D dipolar boson systems for arbitrary dipole tilt angles.
Our results were obtained from detailed calculations at the mean-field level using BdG theory. 
The non-zero tilt angle, as also shown in previous QMC calculations~~\cite{macia2012excitations,macia2014phase}, results in much richer physics for dipolar 2D and quasi-2D bosons,
compared to systems at zero tilt angle. As shown here, the varied behavior at non-zero tilt angles is captured even at the mean-field level,
and our results compare favorably wth QMC calculations (strict 2D regime). Thus, as summarized in our 2D phase diagram (Fig. 4), our calculations demonstrate roton instabilities at large densities for small tilt angles, and at low densities for large tilt angles. The behavior is anisotropic in k-space; accordingly the roton instabilities occur first in the $k_y$ direction, suggestive of inhomogeneity and stripe phase, with density mode becoming soft in the y-direction. Beyond a critical tilt angle, at any density, the dipolar system collapses owing 
to a phonon instability. While our BdG calculations suggest instability at high density, these are not able to obtain 
a quantum solid phase at high densities, as found in QMC calculations. 

There are similarities between the 2D and quasi-2D systems with respect to the excitation spectrum and instabilities. One notable
finding is that the strictly quasi-2D (as opposed to 3D) region is quite small for all tilt angles, and as our quasi-2D phase diagram 
(Fig. 9) shows, at sufficiently large tilt angles, it is possible for a quasi-2D dipole system to be unstable towards a density wave phase.

As pointed out~\cite{ddi-reg}, the inherent issue of ultraviolet divergence 
in the Fourier transform of the dipole interaction in 2D may  be  dealt with in a number of ways; we have chosen to use two possible means of regularization.
For the strict 2D case, our use of a short-range cut-off $r_c$, though based on physical reasoning, does introduce a cut-off dependence in our results,
but the qualitative features are not affected. In the quasi-2D systems we chose, no such cut-off is needed. 
Our calculation of quantum depletion in the quasi-2D case shows that while the BdG treatment may not be as valid near the transition for 
zero tilt angle , it is valid for non-zero tilt angles. For the homogeneous 2D case, quantum depletion depends on the ratio of the dipolar
length $a_{dd}$ and the cut-off $r_c$, becoming smaller for progressively smaller ratio (though the ration needs to be sufficiently large to have a physically
sensible cut-off size).

\section{Acknowledgements}
We would like to thank J. Boronat,  G. Baym, and E. Krotscheck for useful discussions. P. Shen acknowledges support from Institute for Complex Adaptive Matter (ICAM). K. Quader acknowledges the hospitality of Aspen Center for Physics, where part of this work was done.


\bibliography{pengtao_bibtex}

\end{document}